\documentclass{aa}  
\usepackage{multirow}
\usepackage{tablefootnote}

\usepackage{graphicx}
\usepackage{txfonts}
\usepackage{hyperref}
\usepackage{amsmath}
\usepackage{xcolor}
\usepackage{orcidlink}
\usepackage{placeins}

\newcommand{\vox}{\mathrm{vox}}
\newcommand{\kvec}{{\vec{k}}}

\usepackage{array}
\newcolumntype{L}[1]{>{\raggedright\let\newline\\\arraybackslash\hspace{0pt}}b{#1}}
\newcolumntype{C}[1]{>{\centering\let\newline\\\arraybackslash\hspace{0pt}}b{#1}}
\newcolumntype{R}[1]{>{\raggedleft\let\newline\\\arraybackslash\hspace{0pt}}b{#1}}

\newcolumntype{T}[1]{>{\raggedleft\let\newline\\\arraybackslash\hspace{0pt}}p{#1}}
\newcolumntype{Z}[1]{>{\raggedright\let\newline\\\arraybackslash\hspace{0pt}}m{#1}}
\newcolumntype{S}[1]{>{\raggedleft\let\newline\\\arraybackslash\hspace{0pt}}m{#1}}

\begin{document} 
   \title{COMAP Pathfinder -- Season 2 results II.~Updated~constraints~on~the~CO(1--0)~power~spectrum}

    \authorrunning{N.-O.~Stutzer et al.}
    \titlerunning{COMAP Pathfinder -- Season 2 results II.~Updated~constraints~on~the~CO(1--0)~power~spectrum}
   
   \author{
        N.-O.~Stutzer\inst{1}\fnmsep\thanks{\email{\href{mailto:n.o.stutzer@astro.uio.no}{n.o.stutzer@astro.uio.no}}}\orcidlink{0000-0001-5301-1377}
        \and 
        J.~G.~S.~Lunde\inst{1}\orcidlink{0000-0002-7091-8779}
        \and 
        P.~C.~Breysse\inst{6}\fnmsep\inst{7}\orcidlink{0000-0001-8382-5275}
        \and 
        D.~T.~Chung \inst{3}\fnmsep\inst{4}\fnmsep\inst{5}\orcidlink{0000-0003-2618-6504}
        \and 
        K.~A.~Cleary\inst{2}\orcidlink{0000-0002-8214-8265}
        \and 
        D.~A.~Dunne\inst{2}\orcidlink{0000-0002-5223-8315}
        \and 
        H.~K.~Eriksen\inst{1}\orcidlink{0000-0003-2332-5281}
        \and 
        H.~T.~Ihle\inst{1}\orcidlink{0000-0003-3420-7766}
        \and
        H.~Padmanabhan\inst{8}\orcidlink{0000-0002-8800-5740}
        \and
        D.~Tolgay\inst{3}\fnmsep\inst{14}\orcidlink{0000-0002-3155-946X}
        \and
        I.~K.~Wehus\inst{1}\orcidlink{0000-0003-3821-7275}
        \and
        J.~R.~Bond\inst{3}\fnmsep\inst{14}\fnmsep\inst{15}\orcidlink{0000-0003-2358-9949}
        \and 
        S.~E.~Church\inst{17}
        \and 
        T.~Gaier\inst{12}
        \and 
        J.~O.~Gundersen\inst{18}\orcidlink{0000-0002-7524-4355}
        \and 
        A.~I.~Harris\inst{19}\orcidlink{0000-0001-6159-9174}
        \and 
        S.~E.~Harper\inst{10}\orcidlink{0000-0001-7911-5553}
        \and 
        R. Hobbs\inst{9}
        \and 
        J.~Kim\inst{13}\orcidlink{0000-0002-4274-9373}
        \and
        J.~W.~Lamb\inst{9}\orcidlink{0000-0002-5959-1285}
        \and 
        C.~R.~Lawrence\inst{12}\orcidlink{0000-0002-5983-6481}
        \and
        N.~Murray\inst{3}\fnmsep\inst{14}\fnmsep\inst{15}
        \and
        T.~J.~Pearson\inst{2}\orcidlink{0000-0001-5213-6231}
        \and 
        L.~Philip\inst{12}\fnmsep\inst{16}\orcidlink{0000-0001-7612-2379}
        \and
        A.~C.~S.~Readhead\inst{2}\orcidlink{0000-0001-9152-961X}
        \and 
        T.~J.~Rennie\inst{10}\fnmsep\inst{11}\orcidlink{0000-0002-1667-3897}
        \and 
        D.~P.~Woody\inst{9}
        {(COMAP~Collaboration)}
    }

   \institute{
        Institute of Theoretical Astrophysics, University of Oslo, P.O. Box 1029 Blindern, N-0315 Oslo, Norway
        \and 
        California Institute of Technology, 1200 E. California Blvd., Pasadena, CA 91125, USA
        \and 
        Canadian Institute for Theoretical Astrophysics, University of Toronto, 60 St. George Street, Toronto, ON M5S 3H8, Canada
        \and 
        Dunlap Institute for Astronomy and Astrophysics, University of Toronto, 50 St. George Street, Toronto, ON M5S 3H4, Canada
        \and 
        Department of Astronomy, Cornell University, Ithaca, NY 14853, USA
        \and 
        Center for Cosmology and Particle Physics, Department of Physics, New York University, 726 Broadway, New York, NY, 10003, U.S.A 
        \and 
        Department of Physics, Southern Methodist University, Dallas, TX 75275, USA
        \and 
        Departement de Physique Théorique, Universite de Genève, 24 Quai Ernest-Ansermet, CH-1211 Genève 4, Switzerland
        \and 
        Owens Valley Radio Observatory, California Institute of Technology, Big Pine, CA 93513, USA
        \and 
        Jodrell Bank Centre for Astrophysics, Alan Turing Building, Department of Physics and Astronomy, School of Natural Sciences, The University of Manchester, Oxford Road, Manchester, M13 9PL, U.K.
        \and 
        Department of Physics and Astronomy, University of British Columbia, Vancouver, BC Canada V6T 1Z1, Canada
        \and 
        Jet Propulsion Laboratory, California Institute of Technology, 4800 Oak Grove Drive, Pasadena, CA 91109
        \and
        Department of Physics, Korea Advanced Institute of Science and Technology (KAIST), 291 Daehak-ro, Yuseong-gu, Daejeon 34141, Republic of Korea
        \and 
        Department of Physics, University of Toronto, 60 St.~George Street, Toronto, ON, M5S 1A7, Canada
        \and 
        David A.~Dunlap Department of Astronomy, University of Toronto, 50 St.~George Street, Toronto, ON, M5S 3H4, Canada
        \and 
        Brookhaven National Laboratory, Upton, NY 11973-5000
        \and 
        Kavli Institute for Particle Astrophysics and Cosmology and Physics Department, Stanford University, Stanford, CA 94305, USA
        \and 
        Department of Physics, University of Miami, 1320 Campo Sano Avenue, Coral Gables, FL 33146, USA
        \and 
        Department of Astronomy, University of Maryland, College Park, MD 20742
   }
   \date{Received 14 June 2024 / Accepted 30 August 2024}

% \abstract{}{}{}{}{} 
% 5 {} token are mandatory
 
  \abstract
      {We present updated constraints on the cosmological 3D power spectrum of carbon monoxide CO(1--0) emission in the redshift range $2.4$--$3.4$. The constraints are derived from the two first seasons of Carbon monOxide Mapping Array Project (COMAP) Pathfinder line intensity mapping observations aiming to trace star formation during the epoch of galaxy assembly. These results improve on the previous Early Science results through both increased data volume and an improved data processing methodology. On the methodological side, we now perform cross-correlations between groups of detectors (``feed groups''), as opposed to cross-correlations between single feeds, and this new feed group pseudo power spectrum (FGPXS) is constructed to be more robust against systematic effects. In terms of data volume, the effective mapping speed is significantly increased due to an improved observational strategy as well as a better data selection methodology. The updated spherically and field-averaged FGPXS, $\tilde{C}(k)$, is consistent with zero, at a probability-to-exceed of around $34\,\%$, with an excess of $2.7\,\sigma$ in the most sensitive bin. Our power spectrum estimate is about an order of magnitude more sensitive in our six deepest bins across ${0.09\,\mathrm{Mpc}^{-1} < k < 0.73\,\mathrm{Mpc}^{-1}}$, compared to the feed-feed pseudo power spectrum (FPXS) of COMAP ES. Each of these bins individually constrains the CO power spectrum to ${kP_\mathrm{CO}(k)< 2400-4900\,\mathrm{\mu K^2 Mpc^{2}}}$ at $95\,\%$ confidence. To monitor potential contamination from residual systematic effects, we analyzed a set of 312 difference-map null tests and found that these are consistent with the instrumental noise prediction. In sum, these results provide the strongest direct constraints on the cosmological 3D CO(1--0) power spectrum published to date.} %... finally restoring balance to the Force.}

   
   \maketitle

\section{Introduction}
By collecting the combined redshift-dependent line emission from all sources, both diffusely emitting gas and all galaxies, bright and faint, line intensity mapping (LIM) aims to map the Universe from large to small scales in three dimensions \citep[see][and references therein for details on LIM]{madau:1997, battye:2004, peterson:2006, loeb:2008, kovetz:2017, kovetz:2019}. Several emission lines of interest have been proposed, among them $21\,\mathrm{cm}$, carbon monoxide (CO), ionized carbon ([\ion{C}{ii}]), $\mathrm{Ly\alpha}$, and $\mathrm{H\alpha}$, each with different astrophysical and cosmological goals \citep{kovetz:2017,kovetz:2019, bernal:2022}. 

At the forefront of CO LIM is the CO Mapping Array Project (COMAP), currently in its Pathfinder phase, which aims to measure the large-scale CO(1--0) line emission at redshifts of $z\sim2.4$--$3.4$, tracing the star-forming galaxies around the epoch of galaxy assembly \citep{comap-I}. The COMAP Pathfinder instrument is a focal plane array of 19 detectors (which we refer to as "feeds"), each with independent receiver electronics, fielded on a 10.4m Leighton telescope at the Owens Valley Radio Observatory. It observes in a frequency range of $26$--$34\,\mathrm{GHz}$ and is sensitive to $115.27\,\mathrm{GHz}$ CO(1--0) rotational line emission at redshifts of $z\sim 2.4$--$3.4$. Based on the first year of observations (``Season 1''), COMAP obtained the first direct limits on the 3D CO(1--0) clustering power spectrum, already ruling out several models from the literature. These results were published in a series of eight Early Science (ES) papers, along with a preview of our ongoing continuum survey of the Galaxy, a look at the prospects for CO LIM at the epoch of reionization, and a cross-correlation of ES data with an overlapping galaxy survey  \citep{comap-I, comap-II, comap-III, comap-IV, comap-V, comap-VI, comap-VII, comap-VIII}.

In this paper, the second in a series of three, we update our power spectrum results based on observations taken in our first and second seasons (S2), following \citet{comap-IV}. We build on the filtered and calibrated low-level COMAP data products described in detail by {\cite{lunde:2024}}. Implications for astrophysical constraints and modeling are explored by \citet{chung:2024a}.  

As is discussed by \citet{lunde:2024}, the current experimental design is overall very similar to ES, but takes into account a few important lessons learned. For example, COMAP Season 2 uses only constant elevation scans (CES), not Lissajous scans, because one of the main conclusions of \cite{comap-IV} was that changes in elevation within a scan result in significant residual systematic effects from changes in the atmospheric and or ground pickup signals. We also avoid elevations that are strongly contaminated by ground radiation received in the sidelobes. In addition, the instrument drive speed was decreased around May 2022 in order to reduce the stress on the telescope \citep{lunde:2024}, and the effective instrumental properties therefore changed notably about halfway through the second season. We denote periods before and after the speed change the ``fast-'' and ``slow-moving azimuth scans'' respectively (these are equivalent to the naming convention ``Season 1+Season 2a'' and ``Season 2b'' used by \cite{lunde:2024}, where ``a'' and ``b'' denote the period before and after the drive changes). 

For consistency with previous COMAP publications, we adopt the same $\Lambda$CDM cosmological model as \citet{comap-V} and \citet{li:2016} when converting distances in our map cubes from angular and spectral frequency units into physical units. Explicitly we set $\Omega_m = 0.286$, $\Omega_\Lambda = 0.714$, $\Omega_b=0.047$, $H_0=100\,h\,\mathrm{km}\,\mathrm{s}^{-1}\,\mathrm{Mpc}^{-1}$, $h = 0.7$, $\sigma_8 = 0.82$, and $n_s = 0.96$, which is roughly consistent with WMAP \citep{hinsaw:2013}. Unless otherwise stated all distances and distance-derived quantities in megaparsecs carry an implicit $h^{-1}$.

This paper is structured as follows, the power spectrum methodology and updated null test framework are presented in Sect.~\ref{sec:power_spectrum_methodology} and\ref{sec:null_tests_methodology} respectively. In Sect.~\ref{sec:sim_pipeline}, we present the power spectrum transfer function used to account for signal loss from low-level filtering and instrumental effects. Sections ~\ref{sec:power_spectrum_results} and \ref{sec:null_test_results} show the power spectrum results and the outcome of our null tests. Our conclusions are presented in Sect.~\ref{sec:conclusion}.

\begin{table}\caption{\label{tab:saddlebags} Feed groups used in the feed group pseudo-cross-power spectrum.}
    \begin{center}
        \begin{tabular}{p{3cm} p{3cm}}
            \hline \hline
            DCM1 (feed group) & Feeds \\
            \hline
            1\dotfill & 1, 4, 5, 12, 13 \\
            2\dotfill & 6, 14, 15, 16, 17 \\
            3\dotfill & 2, 7, 18, 19 \\
            4\dotfill & 3, 8, 9, 10, 11 \\
            \hline
        \end{tabular}
    \end{center}
    \tablefoot{``Feed groups'' and their associated first down-conversion (DCM1) electronics.}    
\end{table}

\section{Power spectrum methodology}\label{sec:power_spectrum_methodology}
The power spectrum fully characterizes the information contained in a Gaussian random field and so is one of the most powerful statistics for cosmological density fields. While the non-linear physics of galactic emissions to which COMAP is sensitive is not fully Gaussian, the power spectrum is still a useful statistic, and complementary to other summary statistics such as the Voxel Intensity Distribution (VID); \citep{breysse:2017, ihle:2019} or the Deconvolved Distribution Estimator (DDE); \citep{breysse:2023,chung:2023}.

The COMAP Pathfinder uses three-dimensional maps of the CO(1--0) emission to constrain models of star formation during the epoch of galaxy assembly. While the maps already represent the compression of hundreds of terabytes of raw time-ordered data (TOD) into only a few gigabytes, it is possible to encode and compress much of the relevant astrophysical and cosmological information contained within the maps even more by using summary statistics like the power spectrum. As such the power spectra are easier and more computationally efficient to work with when constraining astrophysical and cosmological information of the mapped emission field. 

In the COMAP ES paper series, \cite{comap-IV} devised a novel cross-power spectrum methodology, the feed-feed pseudo-cross-power spectrum (FPXS), constructed to be robust against systematic errors. This work largely builds on the methodology developed by \cite{comap-IV} and lessons learned since the ES data processing to improve the power spectrum constraints of COMAP even further. In the following we summarize the FPXS methodology used and outline what has changed from the methodology developed by \cite{comap-IV}.

\subsection{The Feed Group Pseudo-Cross-Power Spectrum}\label{subsec:FGPXS_summary}
We begin by defining the general concepts of an auto- and cross-power spectrum. The auto-power spectrum can simply be defined as the variance of Fourier modes of a map. It can be written as 
\begin{equation}
    P(\kvec) = \frac{V_\vox}{N_\vox}\langle|\mathcal{F}\{\vec{m}_i\}|^2\rangle = \frac{V_\vox}{N_\vox}\langle|f_i(\kvec)|^2\rangle,\label{eq:auto-spec}
\end{equation}
where $V_\vox$ is the volume of a voxel (i.e., three-dimensional pixel) in units $\mathrm{Mpc}^3$, $N_\vox$ is the number of voxels and $f_i(\kvec)$ are the Fourier coefficients of the map $\vec{m}_i$ at wavenumber $\kvec$. The units of $f_i(\kvec)$ and $\kvec$ are, respectively, the same as the map's $\vec{m}_i$  and $\mathrm{Mpc}^{-1}$. For the Fourier transform $\mathcal{F}\{\vec{m}_i\}$ of the map $\vec{m}_i$ we use the same convention previously used in ES \citep{comap-IV,numpy:2020}. We can safely use the regular Fourier basis in the case of COMAP, instead of the more general spherical harmonics, as the fields are only $\sim 2^\circ$ in diameter and the flat-sky approximation is sufficient. Since our maps are three-dimensional, so is the power spectrum derived from those maps. 

The auto-power spectrum will pick up all components that contribute to the variance in the map: signal, noise, and systematic effects. It can thus be decomposed into 
\begin{equation}
    P(\kvec) = P_\mathrm{CO}(\kvec) + P_\mathrm{noise}(\kvec) + P_\mathrm{syst}(\kvec),
\end{equation}
showing contributions from CO signal\footnote{We note that technically $P_\mathrm{CO}(\kvec)$ in this notation would include contributions from both cosmic CO and all other astrophysical components with non-trivial frequency structure that are not subtracted out by the low-level data analysis steps, e.g., potential interloper line emission. However, for CO(1--0) at $z=2$--$3$ there are very few, if any, interloper lines, apart from a $\sim 10\,\%$ contamination from CO(2--1) at $z=6-8$  \citep{comap-VII, chung:2024b} that could be picked up, and we therefore use a CO-only notation.}, noise and systematic effects, respectively. To obtain an unbiased estimate of the signal power spectrum, the systematic effects and noise properties of the map have to be understood and modeled.

Similarly to the auto-power spectrum, we can define a cross-power spectrum between two maps to be the covariance of Fourier coefficients of the two. It can be written as
\begin{equation}
    C_{ij}(\kvec) = \frac{V_\vox}{N_\vox}\left\langle \operatorname{Re}\{{f_i^*(\kvec) f_j(\kvec)}\}\right\rangle,\label{eq:cross-spec}
\end{equation}
where $f_i$ and $f_j$ represent Fourier coefficients of two different maps $i$ and $j$. The cross-power spectrum reduces to the auto-power spectrum if the two maps $i$ and $j$ are chosen to be identical. 

As opposed to the auto-power spectrum, a cross-power spectrum will only be sensitive to correlated common modes between the two maps. Independent noise and independent systematic effects will therefore be canceled out and we can decompose the cross-spectrum as 
\begin{equation}
    C_\mathrm{ij}(\kvec) = P_\mathrm{CO}(\kvec) + C_{ij,\mathrm{common}}(\kvec),
\end{equation}
where $C_{ij,\mathrm{common}}(\kvec)$ represents the cross-spectrum contribution from some common systematic effect between the two maps. 

We can now see a powerful property of the cross-spectrum: if we can choose two maps with independent noise properties and statistically unique systematic effects, giving $C_{ij,\mathrm{common}}(\kvec) = 0$,  the cross-power spectrum will yield an unbiased estimator for the signal power spectrum $P_\mathrm{CO}(\kvec)$.

This is the main property that the FPXS developed by \cite{comap-IV} is built to exploit. Because the COMAP Pathfinder measures the sky with 19 feeds, each with its own receiver signal chain, the maps from different detectors will have independent noise properties. Additionally, several systematic effects are believed to be unique to each feed, or specific group of feeds. Therefore, a cross-spectrum between two detector maps will not be biased by the noise contribution of the detectors or feed-specific systematic contamination. 

In this work, rather than cross-correlating individual feeds, we instead cross-correlate groups of feeds. In particular, we group feeds by their shared first down-conversion (DCM1) local oscillator \citep{comap-II}. Table~\ref{tab:saddlebags} shows the feeds that are grouped together in a given ``feed group''.

The reason for this change is that some of the systematic effects uncovered with the improved sensitivity of the current data volume are correlated with the DCM1 feed groups as has been shown by \cite{lunde:2024}. Applying the original FPXS, involving cross-correlation of feeds from the same feed group, would not have been effective in canceling such systematic effects since they are common-mode for a given feed group. 

Instead, by grouping all feeds in a given feed group together, detectors from the same feed group are never cross-correlated when computing the average feed group pseudo-cross-power spectra (FGPXS). This effectively cancels the systematic effects that are common to each feed group, while retaining the CO signal. Additionally, grouping together detectors in this way produces effective detector maps that have more sky overlap. Thus, when cross-correlating these maps we obtain better constraints on large-scale power spectrum modes and less mode-mixing due to a larger cross-map footprint. The result from a lower degree of mode mixing is also a lower amount of large-scale systematic effects that can leak into the small- and intermediate-scale power. This is especially important, as we know from \cite{lunde:2024} that our most dominant systematic effects are large-scale modes in the maps.

After splitting the data into feeds or feed groups, we split the data additionally into halves, each with independent systematic effect contributions, such as high or low elevation, as was done by \cite{comap-IV}. This further eliminates unwanted contributions to the cross-spectrum. 

However, even though the FGPXS is slightly more robust to systematic effects, this comes at the price of a slight decrease in sensitivity. The expected loss in sensitivity when using FGPXS as opposed to FPXS should in theory follow the upper limit found by \cite{comap-IV}:
\begin{align}
    \sigma^{N_\mathrm{split}, N_\mathrm{feed}}_{C(\kvec)} &\geq \sqrt{\frac{1}{1 - \frac{1}{N_\mathrm{split}}}}\sqrt{\frac{1}{1 - \frac{1}{N_\mathrm{feed}}}} \sigma_P(k) \label{eq:sensitivity_general}\\
    &\overset{N_\mathrm{split}=2}{=} \sqrt{\frac{2}{1 - \frac{1}{N_\mathrm{feed}}}} \sigma_P(k), \label{eq:sensitivity}
\end{align}
where the uncertainty of a cross-spectrum, $\sigma^{N_\mathrm{split}, N_\mathrm{feed}}_{C(\kvec)}$, is given by the number of cross-correlated data splits and feeds, $N_\mathrm{split}$ and $N_\mathrm{feed}$, compared to the optimal sensitivity, $\sigma_P(k)$, one can obtain when using all available data in an auto-power spectrum. To incorporate the effects of both cross-correlating feeds and an additional cross-variable on the total sensitivity we have generalized Eq. (14) of \cite{comap-IV} to obtain Eq.~(\ref{eq:sensitivity_general}). Because we use the same elevation split as \cite{comap-IV}, splitting the data into high or low elevation, the $N_\mathrm{split}$ dependency of Eq.~(\ref{eq:sensitivity_general}) reduces to the $\sqrt{2}$ loss in sensitivity in Eq.~(\ref{eq:sensitivity}).

\begin{figure}[h!]
    \centering
    \includegraphics[width=\linewidth]{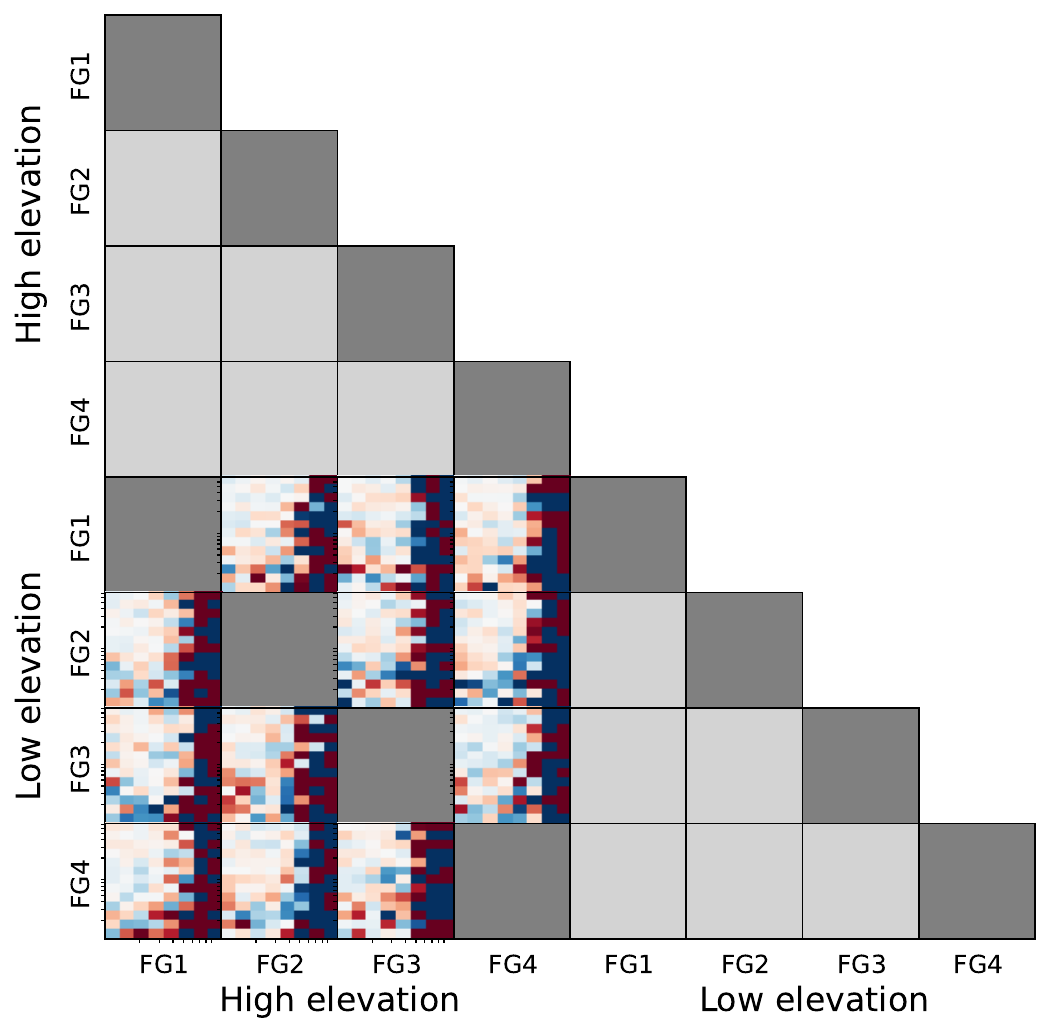}
    \caption{Example grid of possible feed group (FG1--4) and elevation (high and low) split combinations. Combinations with dark and light gray shading, respectively, represent auto-feed group and auto-elevation combinations that are not used in the final averaged FGPXS. The cross-combinations containing examples of a 2D cross-spectrum contribute to the final average FGPXS as neither identical feed groups, nor elevations, are crossed.}
    \label{fig:fg_grid}
\end{figure}

To give some intuition on Eq.~(\ref{eq:sensitivity_general}), we show a grid of possible feed group and elevation split combinations in Fig. \ref{fig:fg_grid}. Equation~(\ref{eq:sensitivity}) can be obtained by using a grid like the one seen in Fig. \ref{fig:fg_grid} from the ratio between the total number of split combinations (i.e the optimal auto-spectrum sensitivity $\sigma_P$) and the number of all cross-combinations that do not constitute auto-combinations between feeds or elevations (respectively dark and light gray shading). From this we should expect there to be a loss in sensitivity in the FGPXS compared to FPXS of $\sim 12\,\%$ -- that is by, respectively, inserting values for the number $N_\mathrm{feed}$ of feed groups, $N_\mathrm{feed} = 4$, and all individual feeds, $N_\mathrm{feed} = 19$, in Eq.~\ref{eq:sensitivity} and comparing the results\footnote{We note that this number in practice tends to be a little larger because we exclude auto-combinations between feed(-groups) and these contain the largest fraction of the optimal total auto-spectrum sensitivity due to better overlapping cross-sky maps.}.

Nevertheless, we conservatively cluster feeds into the aforementioned feed groups to avoid systematic effect contamination, at the price of a minor loss in sensitivity. Apart from the reasons stated above, there is in principle no difference between the FPXS and FGPXS algorithmically; for instance, it would be trivial to group the feeds in a different configuration if that were found to be advantageous in the future for some reason. We can thus describe the two methods using the same algorithmic representation shown in the following. We can write the main steps of the FGPXS algorithmically as follows.
\begin{enumerate}
    \item Split the data into two halves $A$ and $B$. As was done by \cite{comap-IV}, we chose elevation as the main cross-correlation variable to eliminate potential sidelobe pickup from the ground.
    \item For parts $A$ and $B$ respectively make maps of each feed group $i$. We denote these by, for example, $m_{A_2}$ for a map of part $A$ with feed group $2$.
    \item For each combination of feed groups $i$ and $j$, and data splits $A$ and $B$,  compute cross-power spectra.
    \item Compute a noise-weighted average FGPXS of all the resulting $N_\mathrm{feed group}\times (N_\mathrm{feed group} - 1)$ (with $N_\mathrm{feed group} = 4$ when using feed groups and $N_\mathrm{feed group} = 19$ if computing the ES FPXS) individual FGPXS that do not involve the same detector or elevation:
        \begin{equation}
            C(\kvec) = \frac{\sum_{A_i \neq B_j} \frac{C_{A_i B_j}(\kvec)}{\sigma^2_{C_{A_i B_j}}(\kvec)}}{\sum_{A_i \neq B_j} \frac{1}{\sigma^2_{C_{A_i B_j}}(\kvec)}},\label{eq:coadd_FGPXS}
        \end{equation}
    with a corresponding uncertainty of 
        \begin{equation}
            \sigma_{C(\kvec)} = \frac{1}{\sqrt{\sum_{A_i \neq B_j} \frac{1}{\sigma^2_{C_{A_i B_j}}(\kvec)}}}.\label{eq:coadd_error_FGPXS}
        \end{equation}
    This is what we refer to as the mean feed group pseudo-cross-power
spectrum (FGPXS) or feed-feed pseudo-cross-power
spectrum (FPXS), if (respectively) feed groups or feeds are used as effective detectors. 
\end{enumerate}

In Fig. \ref{fig:fg_grid} we illustrate a grid of possible feed group and elevation combinations used for an average FGPXS. Those shaded dark and light gray represent auto-feed and auto-elevation combinations (respectively). The combinations that cross neither feed groups nor elevations, indicated with examples of 2D FGPXS combinations, are used in the final average FGPXS in Eq.~(\ref{eq:coadd_FGPXS}). 

Due to the non-uniform coverage of our sky fields, as well as a non-trivial survey footprint \citep[see ][for examples of maps]{lunde:2024}, the maps are weighted prior to computing their Fourier coefficients. We use the same weighting scheme as \cite{comap-IV}, given for a cross-power spectrum by
\begin{equation}
    \vec{w}_{A_iB_j} \propto \frac{1}{\vec{\sigma}_{A_i} \vec{\sigma}_{B_j}},
\end{equation}
where $\vec{\sigma}_{Y_x}$ represents the uncertainty estimate in each voxel of a feed group and elevation split map, $\vec{m}_{Y_x}$. These weights are then applied to the map, $\tilde{\vec{m}}_i = \vec{w}_i \vec{m}_i$, before power spectrum estimation with the Fourier coefficients $\tilde{f}_i(\kvec) = \mathcal{F}\{\vec{w}_i \vec{m}_i\}$ in Eq.~(\ref{eq:cross-spec}). Regions outside the map footprints are assigned zero weights. The power spectra of these weighted maps are commonly referred to as pseudo power spectra \citep{hivon:2002}. The pseudo power spectra are a biased power spectrum estimator because different Fourier modes become coupled via the applied weights \citep[see][for details on mode mixing]{hivon:2002,leung:2022}. It should be noted that we use $\tilde{P}(k)$ to denote pseudo spectra in the later results sections, but we use the notation $P(k)$ (without the tilde) in the methods sections as most of the methodology is equivalently written for unbiased and pseudo spectra. A detailed discussion of the COMAP-specific mode mixing can be found in Fig. 1 and Appendix D of \cite{comap-IV}, which shows that the effect is $\leq 20\,\%$ over our k-range. Reversing the mode-mixing will thus be left as a future exercise and is beyond the scope of this work.  

\subsection{The binned power spectrum estimator}\label{subsec:binning}
As COMAP produces line intensity maps spanning three-dimensional redshift-space volumes, the resulting power spectra also span three-dimensional Fourier-space volumes. It can, however, be easier to work with a power spectrum spherically averaged down to one dimension. For the spherically averaged power spectrum to contain all relevant information in the full three-dimensional power spectrum, the emission field is technically required to remain statistically isotropic on large scales and stationary across the mapped redshift range. This is not strictly known to be true for the CO emission field. Cosmic star-formation, especially dust-obscured star-formation history traced in the infrared, is poorly constrained in our targeted redshift range of $z=2.4$--$3.4$ \citep[see][for a review of cosmic star-formation history]{Madau:2014}. As a consequence, the extent to which the mapped CO emission field is stationary is largely unknown. The spherically averaged power spectrum of a dynamic field will not be sensitive to changes in the CO emission across cosmic time, but it will measure the time-averaged properties of the targeted CO structures. However, the distinction is moot for the current COMAP signal-to-noise ratio (S/N), as no clear CO excess is observed in the power spectra. Thus we present the spherically averaged, 1D power spectrum as our main science product.

Additionally, in practice, the angular and the redshift axes are observed in fundamentally different ways, and the low-level filtering applied to the data \citep{lunde:2024} as well as redshift space distortions and line-broadening \citep{chung:2021} can affect the signal and sensitivity differently along each axis. Therefore, the angular and line-of-sight dimensions are convenient to separate, and we bin the 3D power spectrum $C(\kvec)$, with $\kvec = (k_x, k_y, k_z)$, into both a cylindrical and spherically averaged power spectrum. The former of these conserves the structures perpendicular and parallel to the line-of-sight by only merging the two angular axes
\begin{equation}
    \kvec_i=(k_\perp, k_\parallel)=\left(\sqrt{k_x^2 + k_y^2}, k_z\right).
\end{equation}
Meanwhile, the latter will average the 3D power spectrum into 1D bins of the form 
\begin{equation}
    k_i=\sqrt{k_x^2 + k_y^2 + k_z^2}.
\end{equation}

The binned cylindrically averaged power spectrum estimator will then become
\begin{equation}
    C(\kvec) \approx C_{\kvec_i} = \frac{V_\vox}{N_\vox N_\mathrm{modes}}\sum_{\kvec\in \kvec_i}\left\langle \operatorname{Re}\{{f_1^*(\kvec) f_2(\kvec)}\}\right\rangle,
\end{equation}
where the number of Fourier modes in bin $\kvec_i$ is given as $N_\mathrm{modes}$. The equation is completely analogous when binning to the spherically averaged power spectrum. We henceforth refer to the cylindrically and spherically averaged power spectrum estimators as ``2D'' and ``1D'' due to the number of axes needed to display them, but note that they still represent averages of a 3D density field. 

The bin edges are chosen to cover the scales to which COMAP is most sensitive and correspond to those used in the COMAP ES power spectra \citep{comap-IV}, but due to our large increase in sensitivity and better understanding of the origin of correlations on large angular scales we conservatively excise all perpendicular scales $k_\perp \lesssim 0.1\,\mathrm{Mpc}^{-1}$ for this publication. On these scales, we are dominated by sub-optimal cross-map overlap that results in poor constraining power of the large-scale structure as well as the possibility of large-scale residual systematic effect leakage through mode mixing into the small-scale power spectrum modes \citep[see][for examples of our map-domain systematic effects]{lunde:2024}. In the future, we aim to recover the large scales at $k_\perp \lesssim 0.1\,\mathrm{Mpc}^{-1}$. Lastly, we also mask the bins corresponding to the highest $k_\perp$ and $k_\parallel$ used by \cite{comap-IV} to prevent issues with aliasing near the Nyquist frequency of the two respective dimensions: $k_\perp^\mathrm{Nyquist}\approx1.22\,\mathrm{Mpc}^{-1}$ and $k_\parallel^\mathrm{Nyquist}\approx0.74\,\mathrm{Mpc}^{-1}$.

\subsection{Uncertainty estimation from randomized null maps}\label{subsec:rnd_errors}
In order to compute the mean FGPXS and its errors, as is shown in Eqs.~(\ref{eq:coadd_FGPXS}) and (\ref{eq:coadd_error_FGPXS}), we need the power spectrum uncertainties for each feed group and elevation cross-combination, $\sigma^2_{C_{A_i B_j}}(\kvec)$. This can be done via two basic approaches: simulations and data-driven methods. Here, we first detail some problems with a simulation-based approach used previously in COMAP ES \citep{comap-IV} and subsequently argue for why a data-driven approach was chosen in this work.

In COMAP ES, \cite{comap-IV} chose a simple simulation approach where the power spectrum uncertainties were computed from an ensemble of simple white noise maps, ${\vec{m}_{\mathrm{noise}, i} \sim \mathcal{N}(0, \vec{\sigma})}$, drawn from a zero-mean Gaussian distribution $\mathcal{N}$ with the voxel uncertainties $\vec{\sigma}$. These were then propagated to the power spectrum level. The main advantage of this approach is its computational efficiency. However, it can only reflect the white noise level within the map, while residual correlated noise and the effect of the pipeline filters on the noise will not be contained in the uncertainty from these simple white noise maps \citep[see, for instance, the power spectral density (PSD) of TOD in Fig. 9 of][for an illustration of the noise properties of the filtered data]{lunde:2024}. The simplified simulations proved an adequate method given the sensitivity of our ES data. With the increased sensitivity achieved at the end of S2, obtaining suitable power spectrum errors, $\sigma_{C_\kvec}$, through simulations would require the sampling of noise from the TOD domain (ideally with additional ground-up modeling of all contributing systematic effects), propagating it all the way through the low-level pipeline \citep{lunde:2024} up to the power spectrum. However, this would be computationally expensive because the low-level pipeline filters would have to be re-run for each ensemble, and require significant additional data modeling. 

Given the drawbacks of both the white noise and a potential TOD-level simulation-based approach, a data-driven method was instead chosen for this work as it represents a relatively computationally inexpensive method of estimating the power spectrum uncertainties that automatically reflects all the properties of the data. In particular, we draw from the simple idea that we can cancel the signal and systematic effects in a subtraction between data-half maps while leaving the correct noise properties. In our case, we estimate $\sigma_{C_\kvec}$ by what we refer to as an ensemble of randomized null difference (RND) maps.

The first step in the RND calculation is to divide the set of all scans in the data into two randomized halves, $A$ and $B$, from which we subsequently make maps $\vec{m}^{\mathrm{RND}}_{A,i}$ and $\vec{m}^{\mathrm{RND}}_{B,i}$. This is done for all random split realizations $i$. Both $\vec{m}^{\mathrm{RND}}_{A,i}$ and $\vec{m}^{\mathrm{RND}}_{B,i}$ should contain the same signal, and due to the randomization of the splits also the same systematic effects. Hence we can cancel both the signal and systematic effects by computing the difference between the two maps;
\begin{equation}
    \Delta\vec{m}^{\mathrm{RND}}_i = \frac{\vec{m}^{\mathrm{RND}}_{A,i} - \vec{m}^{\mathrm{RND}}_{B,i}}{2}.
\end{equation}
The difference maps $\Delta\vec{m}^{\mathrm{RND}}_i$ now optimally capture the white and correlated noise properties and biases (from low-level filters, the instrumental beam, etc.) of the real maps, but are without any of the signal or systematic effects. As such they reflect the true properties of the data to a high degree.

Finally, to obtain the uncertainty of the power spectrum $\sigma_{C_\kvec}$ we need to compute the FGPXS of each difference map $\Delta\vec{m}^{\mathrm{RND}}_i$. From the resulting ensemble of such feed group cross-spectra, $C_{\Delta m_i}^\mathrm{RND}(\kvec)$, we can compute the uncertainties $\sigma_\mathrm{RND}(\kvec)$ by taking the standard deviation over the ensemble. These can then be used when co-adding together feed group spectra to obtain the final mean FGPXS as is explained in Eqs.~(\ref{eq:coadd_FGPXS}) and (\ref{eq:coadd_error_FGPXS}).

\section{Power spectrum null tests}\label{sec:null_tests_methodology}
With the increased effective COMAP data volume and the resulting increased sensitivity comes the need for more effective null tests to ensure the data quality of our final power spectra.  

As we explain in this section, the null tests devised in this work are based on difference maps in a similar way to the RND method used for uncertainty estimation described earlier in Sect.~\ref{subsec:rnd_errors}, except we are now splitting the maps on meaningful parameters instead of randomly. The goal then becomes finding null variables (e.g. high or low humidity or left or right moving scans; see Table \ref{tab:null_variables} for a list of all chosen variables) that correlate to systematic effects in one of the null variable halves by which we split the data. 

We can write the difference map of some null variable $j$ as 
\begin{equation}
    \Delta \vec{m}^\mathrm{null}_j = \frac{\vec{m}_{A, j} - \vec{m}_{B, j}}{2},\label{eq:diff_null}
\end{equation}
where the maps $\vec{m}_{A, j}$ and $\vec{m}_{B, j}$ represent the maps of the two halves of the data respectively. If the chosen null variable correlates to a systematic effect, the difference map $\Delta \vec{m}^\mathrm{null}_j$ will contain the systematic effect but cancel the signal. The difference maps can then be used to perform a null test, with the null hypothesis being that the null maps are consistent with the general noise properties of the maps. The associated voxel uncertainty of the null map is then given by 
\begin{equation}
    \vec{\sigma}_{\Delta \vec{m}_j}^\mathrm{null} = \frac{\sqrt{\vec{\sigma}_{m_{A, j}}^2 + \vec{\sigma}_{m_{B, j}}^2}}{2},\label{eq:null_sensitivity}
\end{equation}
for uncertainties $\vec{\sigma}_{m_{A, j}}$ and $\vec{\sigma}_{m_{B, j}}$ of the maps $\vec{m}_{A, j}$ and $\vec{m}_{B, j}$ respectively. 

For each of the null variables $j$ we then take the difference between the two maps as is described by Eqs.~(\ref{eq:diff_null}) and (\ref{eq:null_sensitivity}). As we use a cross-elevation FGPXS we must compute a difference map for both high and low elevation. The data are therefore split into four parts, two elevation ranges and two null variables halves, where we subtract across the latter in the map domain and cross-correlate the resulting null maps across the former using the FGPXS method described earlier. With the set of resulting null test FGPXS $C_{\Delta m_j}^{\kvec_i}$ we can perform a $\chi^2$-test, with a null hypothesis that the difference maps are consistent with noise, by first computing 
\begin{equation}
    \chi^2_{\mathrm{null}, j} = \sum_{\kvec_i} \left(\frac{C_{\Delta m_j}^{\kvec_i} - \mu_{\Delta m_j}^{\kvec_i}}{\sigma_{C_{\Delta m_j}^{\kvec_i}}}\right)^2 = \sum_{\kvec_i} \left(\frac{C_{\Delta m_j}^{\kvec_i}}{\sigma_{C_{\Delta m_j}^{\kvec_i}}}\right)^2,
\end{equation}
with the expectation value of the null FGPXS $\mu_{\Delta m_j}^{\kvec_i}= 0$ under the null hypothesis. Here $\sigma_{C_{\Delta m_j}^{\kvec_i}}$ is the uncertainty of the null FGPXS $C_{\Delta m_j}^{\kvec_i}$ in bin $k_i$ for null variable $j$ that is estimated using the RND method described earlier in Sect.~\ref{subsec:rnd_errors}. 

Thereafter we can compute the probability-to-exceed (PTE) that quantifies the probability of obtaining a value $\chi^2_{\mathrm{null}, j}$ or higher. The PTE is defined as 
\begin{equation}
    \mathrm{PTE}(\chi^2) = 1 - \mathrm{CDF}(\chi^2), 
\end{equation}
where for a given probability distribution function $P(\chi^2)$ of the $\chi^2_{\mathrm{null}, j}$ values the corresponding cumulative distribution function is denoted as $\mathrm{CDF}(\chi^2)$. 

In our case, $P(\chi^2)$ does not follow the usual analytical $\chi^2$-distribution because the noise properties of the FGPXS are not completely known analytically \citep[see][for some examples of how cross-spectrum noise properties can look]{watts:2020, nadarajah:2016, gaunt:2019}. We thus compute the PTEs numerically by using an ensemble of RND maps equivalent to those we already use for estimating uncertainties as these will perfectly reflect the noise properties and biases in the data, as well as obey the null hypothesis. For each separate processing run -- over fields, fast- and slow-moving azimuth data -- we compute 244 RND maps, of which we use 61 for power spectrum uncertainty estimation and the remaining 183 for measuring the numerical $\chi^2$-distribution. 

\section{Transfer functions}\label{sec:sim_pipeline}
As is described by \cite{comap-III} and \cite{comap-IV} the COMAP maps are not unbiased as the low-level filtering of the data, the binning of the data into voxels, and the finite resolution of the telescope beam will attenuate the signal in the maps. In this section, we explain how we de-bias our power spectrum estimates using transfer functions for each of the main effects that result in signal loss. The beam and voxel window smoothing of the signal is corrected using analytically computed transfer functions, while the low-level filtering attenuation is quantified using simulations. Details on how each transfer function is estimated are discussed by \cite{lunde:2024}. 

\begin{figure}
    \centering
    \includegraphics[width=\linewidth]{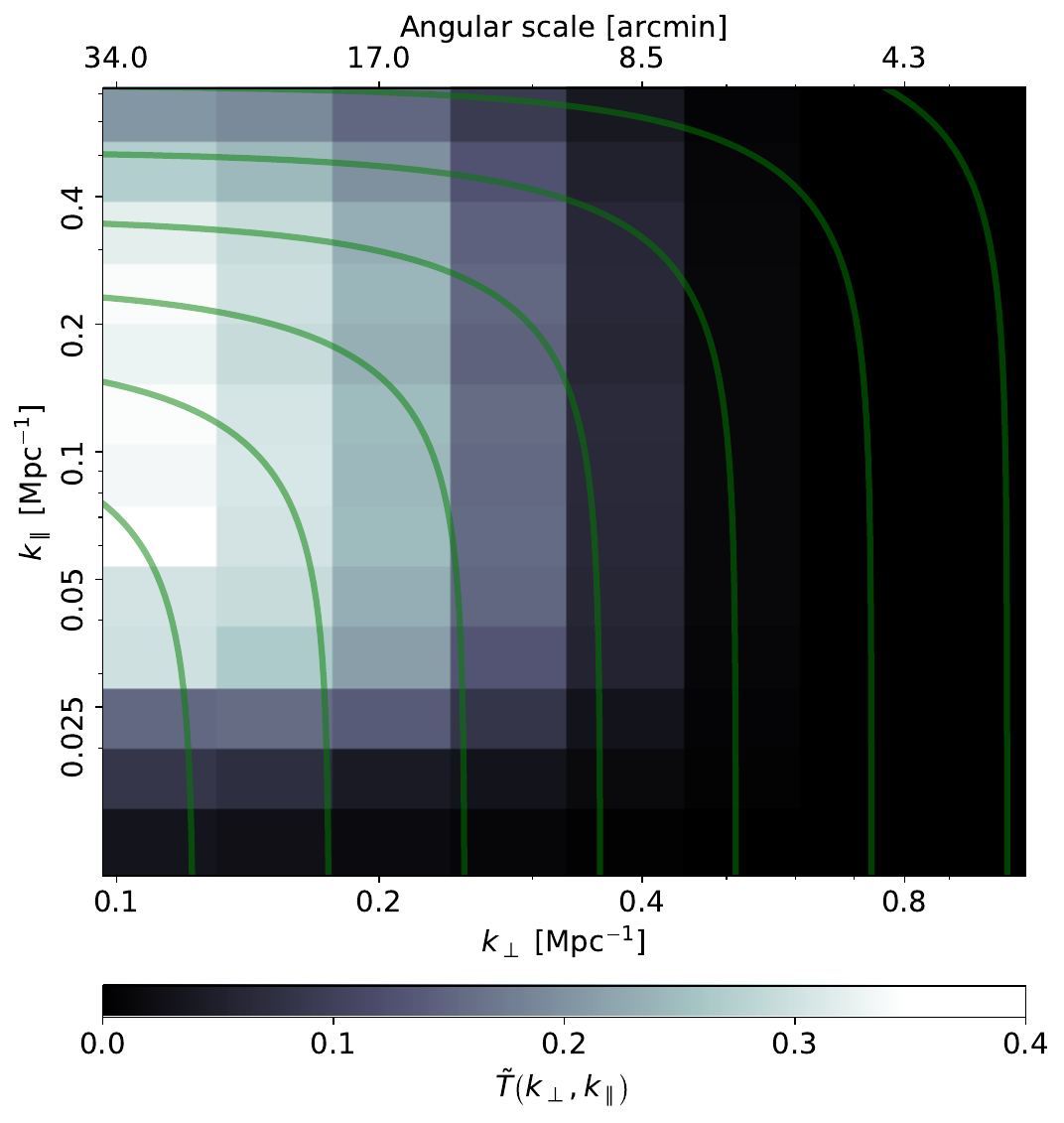}
    \caption{Full power spectrum transfer function used to account for the signal losses due to the low-level filtering pipeline, the instrumental beam smoothing, and the voxel window of the maps \citep[see][for details on each individual transfer function]{lunde:2024}. Thin green contours indicate the bin edges of the (1D) spherically averaged FGPXS.}
    \label{fig:transfer_function}
\end{figure}

When performing a power spectrum analysis of our maps, as is described in Sect.~\ref{sec:power_spectrum_methodology}, we obtain an estimate of the signal that is biased by several different effects. To see how the signal is biased we can write the FGPXS signal estimator as 
\begin{equation}
    C_\kvec = T(\kvec) P^\mathrm{CO}_\kvec = T_f(\kvec)T_\mathrm{b}(k_\perp)T_p(k_\perp)T_\nu(k_\parallel)P^\mathrm{CO}_\kvec,
\end{equation}
where the transfer function $T(\kvec)$ is the product of the filter transfer function $T_f(\kvec)$, the beam smoothing transfer function $T_\mathrm{b}(k_\perp)$ as well as the pixel and spectral channel windows, $T_p(k_\perp)$ and $T_\nu(k_\parallel)$. The transfer function can be written in this multiplicative form in the Fourier domain because the low-level filtering and the smoothing of small-scale structures due to the instrumental beam and voxel window of the map grid can all (approximately) be expressed as a convolution in map domain. In Fig.~\ref{fig:transfer_function} we show the full transfer function product $T(\kvec)$, while the individual transfer function elements are shown in detail in Sect.~6. of \cite{lunde:2024}.

Using our transfer function estimate we can de-bias the FGPXS by deconvolution;
\begin{equation}
    P^\mathrm{CO}_\kvec = \frac{C_\kvec}{T(\kvec)}, 
\end{equation}
with the uncertainties of the signal estimator being affected in a similar manner,
\begin{equation}
    \sigma_{P_\kvec}^\mathrm{CO} = \frac{\sigma_{C_\kvec}}{T(\kvec)},
\end{equation}
becoming large whenever the transfer function $T(\kvec)$ becomes small.

\begin{figure*}
    \centering
    \includegraphics[width=\linewidth]{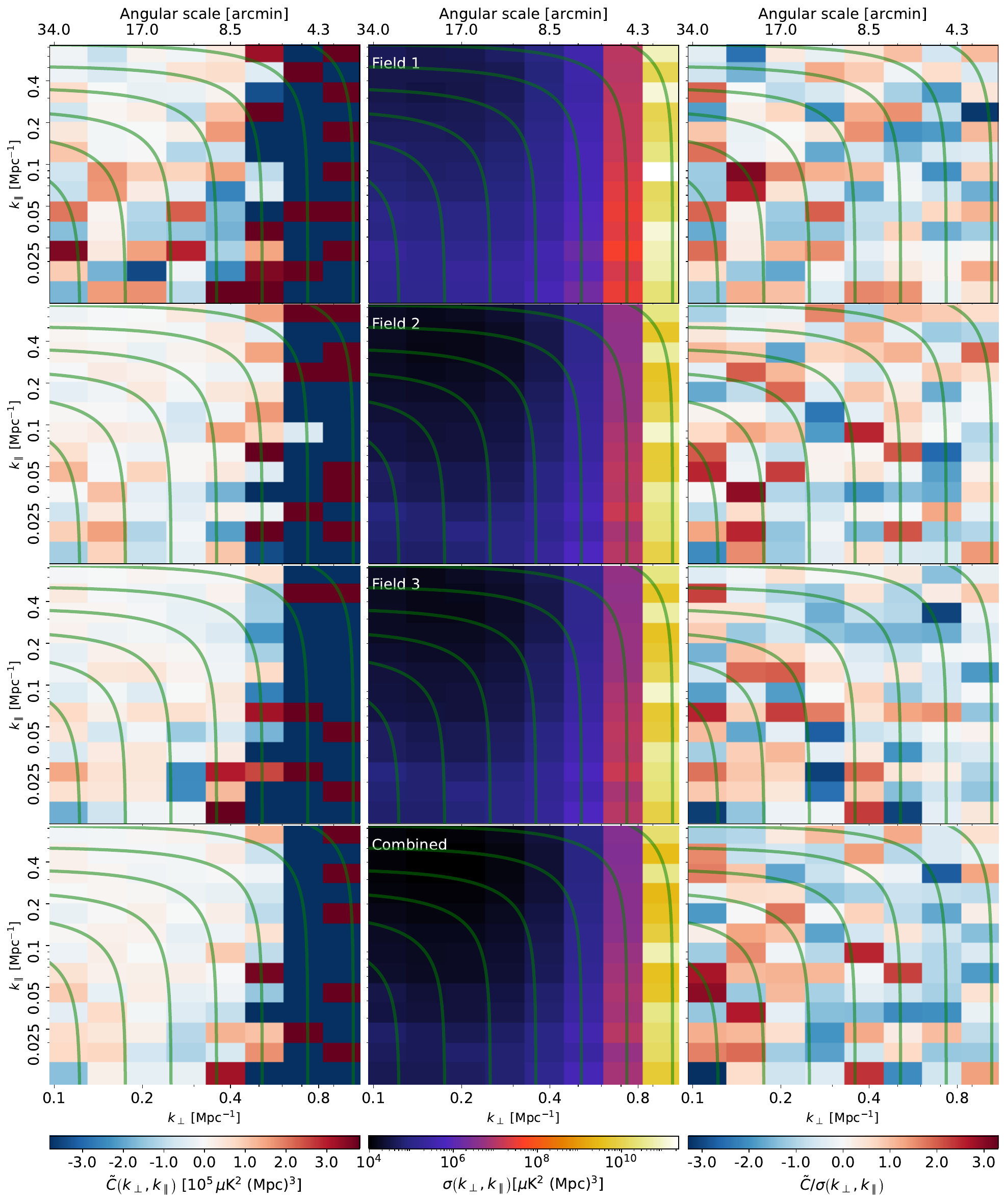}
    \caption{Cylindrically averaged (2D) feed group pseudo-cross-spectra. Columns show, from left to right, the full spectra, the corresponding $1\sigma$ uncertainty, and the ratio between the two. Rows show, from top to bottom, Fields 1, 2, 3, and all three combined. The approximate angular scale, assuming the central COMAP redshift at $z = 2.9$, corresponding to each $k_\bot$ is shown as a twin axis on the upper row of plots. Thin green contours indicate the bin edges of the (1D) spherically averaged FGPXS.}
    \label{fig:pxs_sdlb_2d}
\end{figure*}

\begin{figure}[h!]
    \centering
    \includegraphics[width=\linewidth]{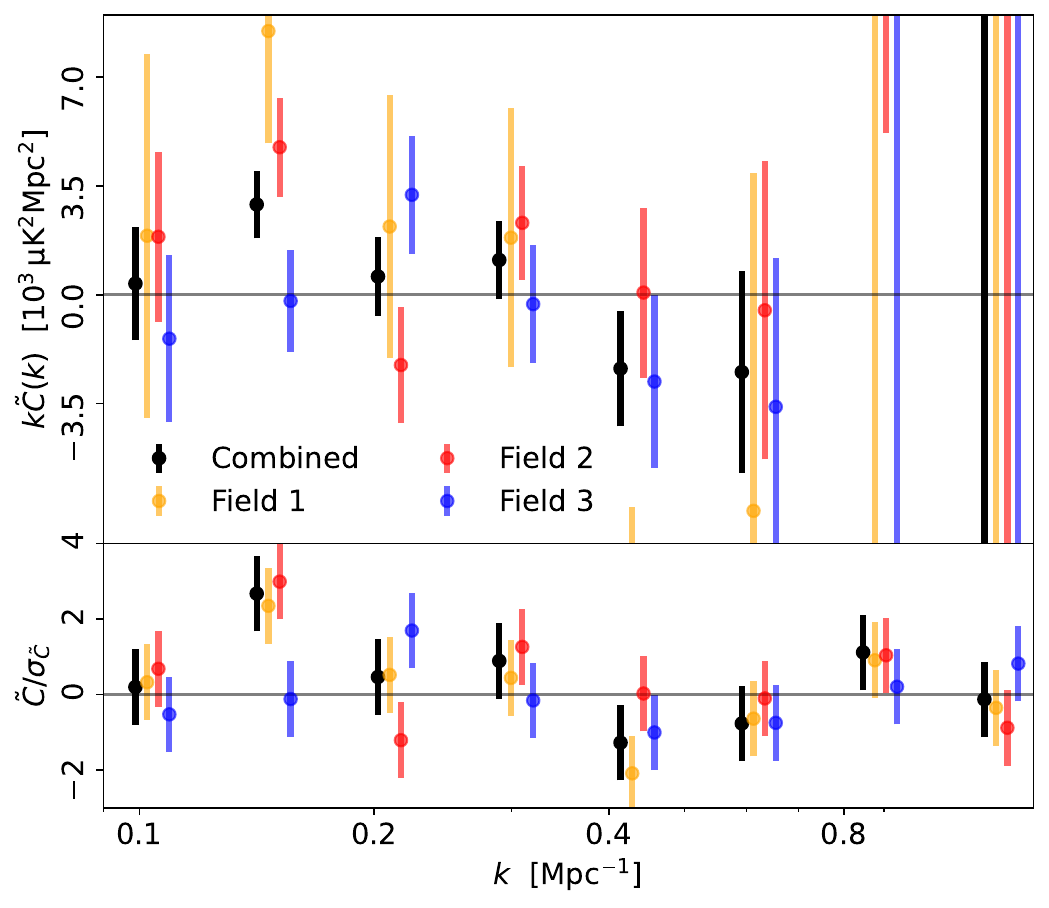}
    \caption{Spherically averaged FGPXS with $1\sigma$ uncertainties for Fields 1, 2 and 3 (orange, red and blue respectively) as well as the combination of the three (black) in units $\mathrm{\mu K^2Mpc^2}$ (\textit{upper panel}) and in units of the $1\sigma$ power spectrum uncertainties (\textit{lower panel}). The data points have been slightly offset from their true $k$-position to increase readability (see Table \ref{tab:power_spectrum values} for an overview over bin centers, FGPXS values, and uncertainties).}
    \label{fig:pxs_1d_per_field}
\end{figure}

\section{Power spectrum results}\label{sec:power_spectrum_results}
In this section, we present the main power spectrum results of this paper. The raw data going into the power spectra are filtered, calibrated, and binned into maps after a set of data selection steps that remove scans that are likely contaminated by systematic effects. This is described in detail by \cite{lunde:2024}.

As one of the main lessons learned from COMAP ES was to employ only CES scans, and no longer use a Lissajous scanning strategy, the data presented here consist only of CES data \citep{comap-III, comap-IV}. Specifically, we include all data obtained up to November 2023, both the ES (Season 1) CES data as well as all data gathered in S2. The data volume obtained in S2 is, as has been explained by \cite{lunde:2024}, effectively around eight times larger than the Season 1 CES data after data selection. In addition, the ES analysis of Season 1 data excluded several detectors that were either offline or excluded due to clear signs of systematic excess in reduced $\chi^2$-tests or in visual inspections of feed cross-spectra. We are able to include these in the S2 analysis because all feeds were functioning during S2 and the map-domain PCA described in \cite{lunde:2024} strongly suppresses detector-specific systematic effects. 

We note that in the ES analysis \cite{comap-IV} removed feed-feed cross-spectra both through a reduced $\chi^2$-test, and manual inspection of misbehaving feed combinations. Due to better low-level data processing, we were able to remove all data-driven cuts in the power spectrum domain, with the increased set of null tests since ES working as an additional safeguard against systematic effects (see Sect.~(\ref{sec:null_test_results}) for discussion of null test results). 

Lastly, in Appendix \ref{apdx:signal_inj_result}, we show a simple end-to-end signal injection test as a qualitative test of our pipeline's ability to recover a known signal's amplitude within the estimated experimental errors and power spectrum transfer function.

\subsection{The cylindrically averaged power spectrum result}\label{subsec:cyl_avg_ps}
In Fig.~\ref{fig:pxs_sdlb_2d} we show the cylindrically averaged (2D) mean FGPXS for all three fields separately, as well as in combination. The figure also shows the sensitivity per $(k_\perp, k_\parallel)$-bin as well as the FGPXS in units of the sensitivity. 

When looking at the 2D FGPXS in Fig.~\ref{fig:pxs_sdlb_2d} we note that the noise blows up on small scales, particularly so in the angular direction, due to the COMAP transfer function seen in Fig.~\ref{fig:transfer_function} \citep[][]{lunde:2024}. However, we see no obvious patterns in the 2D FGPXS that would indicate a systematic effect. In fact, the spectra resemble white noise.

As was mentioned earlier, in Sect.~\ref{subsec:binning}, we want to avoid issues with poorly constrained large-scale modes,  strong mode mixing, and possible residual large-scale systematic effects. We mitigate these issues by excluding 2D bins at $k_\perp < 0.1\,\mathrm{Mpc}^{-1}$. An example of spurious fluctuations induced by poor overlap can be seen in the COMAP ES cylindrically averaged FPXS of Field 1 \citep[see][noting that Field 1 is especially susceptible to poor detector overlap due to its position at declination zero]{comap-IV} as correlated structures along constant $k_\perp$ at scales $k_\perp < 0.1\,\mathrm{Mpc}^{-1}$. These correlations have since been understood to originate from sub-optimal detector overlap, and are pushed to larger scales due to a larger sky overlap when computing cross-spectra between feed groups instead of individual feeds. In interim estimates we found the average of the maximum correlations between a bin and all the others to be around $15\,\%$ at scales $k_\perp \geq 0.1\,\mathrm{Mpc}^{-1}$, while the correlations at $k_\perp \leq 0.1\,\mathrm{Mpc}^{-1}$ are somewhere in the $30$--$70\,\%$ regime. Improved modeling of these correlations will be the aim of future work.

\subsection{The spherically averaged power spectrum result}\label{subsec:sph_avg_ps}
As interpreting the 2D cylindrically averaged FGPXS can be somewhat unintuitive we can bin the spectra into 1D by performing a full spherical averaging. This is done as is described in Sect.~\ref{subsec:binning}, in which the 1D bin-edges are indicated as thin green contours in Fig.~\ref{fig:pxs_sdlb_2d}. When doing so we obtain the spherically averaged FGPXS for the three fields, as well as the combination thereof, as is seen in Fig.~\ref{fig:pxs_1d_per_field}. 

As was discussed in Sect.~\ref{subsec:cyl_avg_ps}, we excluded scales $k_\perp < 0.1\,\mathrm{Mpc}^{-1}$ from the power spectrum analysis to avoid issues with poor cross-map overlap, mode mixing and large correlations between large scale bins. Therefore, Fig.~\ref{fig:pxs_1d_per_field} only shows FGPXS data points on scales $k > 0.1\,\mathrm{Mpc}^{-1}$. Similar to the discussion in Sect.~\ref{subsec:cyl_avg_ps}, we estimate the average of the maximum correlation between a 1D bin and all the others, on scales $k > 0.1\,\mathrm{Mpc}^{-1}$, to be $\lesssim 10\,\%$ after excluding the large $k_\perp$ scales and performing the spherical averaging. Given this $\lesssim 10\,\%$ level, we shall assume for Season 2 analyses downstream that the spherically averaged 1D FPGXS bins are approximately uncorrelated. As with the 2D FGPXS discussed in Sect.~\ref{subsec:cyl_avg_ps}, we intend to improve the exact modeling of these correlations in future work.

When looking at Fig.~\ref{fig:pxs_1d_per_field} we note that Fields 2 and 3 have the highest sensitivity, while Field 1 has around $50\,\%$ larger errors than the two other fields. This is because, of the three COMAP fields, Field 1 is most affected by the low-level data cuts \citep[see][]{lunde:2024}, and due to its location at zero declination, is particularly susceptible to poor detector overlap. In addition, by rejecting poorly overlapping detector combinations, which are also the least sensitive, we prevent the resulting strong mode mixing and potential consequent leakage of systematic effects into smaller scales at the cost of a relatively minimal loss in sensitivity. As a consequence, we keep $2/3$ of the feed group cross-spectra of Field 1 in the analysis.

As we can see from Fig.~\ref{fig:pxs_1d_per_field}, the cross-spectra are largely consistent with zero to within $2\sigma$ in most bins. However, perhaps the most notable feature is the high power in the second and most sensitive $k$-bin, at $0.12\,\mathrm{Mpc}^{-1} < k < 0.18\,\mathrm{Mpc}^{-1}$, which is respectively at around $2.3\sigma$ and $3\sigma$ significance above zero for Fields 1 and 2. Meanwhile, for Field 3 the same bin is consistent with zero power. When combining the three fields, the co-added data point in the second $k$-bin has a value that is $2.7\sigma$ away from zero. For each of the spherically averaged FGPXS of the three fields, we compute their $\chi^2$ PTE to check their constancy with zero power. In doing so, we obtain PTEs of, respectively, $33.2\,\%$, $19.5\,\%$, and $82.7\,\%$ for Fields 1--3. The field-combined spherically averaged FGPXS results in a $34\,\%$  PTE. As for the null tests, the PTE is estimated from the numerical RND $\chi^2$ ensemble. While the combined 1D $\sim 2.7\sigma$ power in $0.12\,\mathrm{Mpc}^{-1} < k < 0.18\,\mathrm{Mpc}^{-1}$ bin is interesting, we do not consider it a statistically significant excess given the estimated PTEs. Thus, we shall have to wait for future analyses, and more data, to answer definitively whether this excess is simply noise or not. 

Although we do not consider the field-combined $2.7\sigma$ point statistically significant, the agreement between two of the three fields is interesting to note. As Fields 1 and 2 are quite different in terms of their path across the sky as it is seen from the telescope (Field 1 being a zero declination field while Field 2 is at a declination of $52.30^\circ$; see \cite{comap-III} for details on the fields) they also are expected to have some independent systematic effects. However, Fields 2 and 3 are more alike and would be expected to share certain systematic effects. 

\begin{figure*}[!ht]
    \centering
    \includegraphics[width=\linewidth]{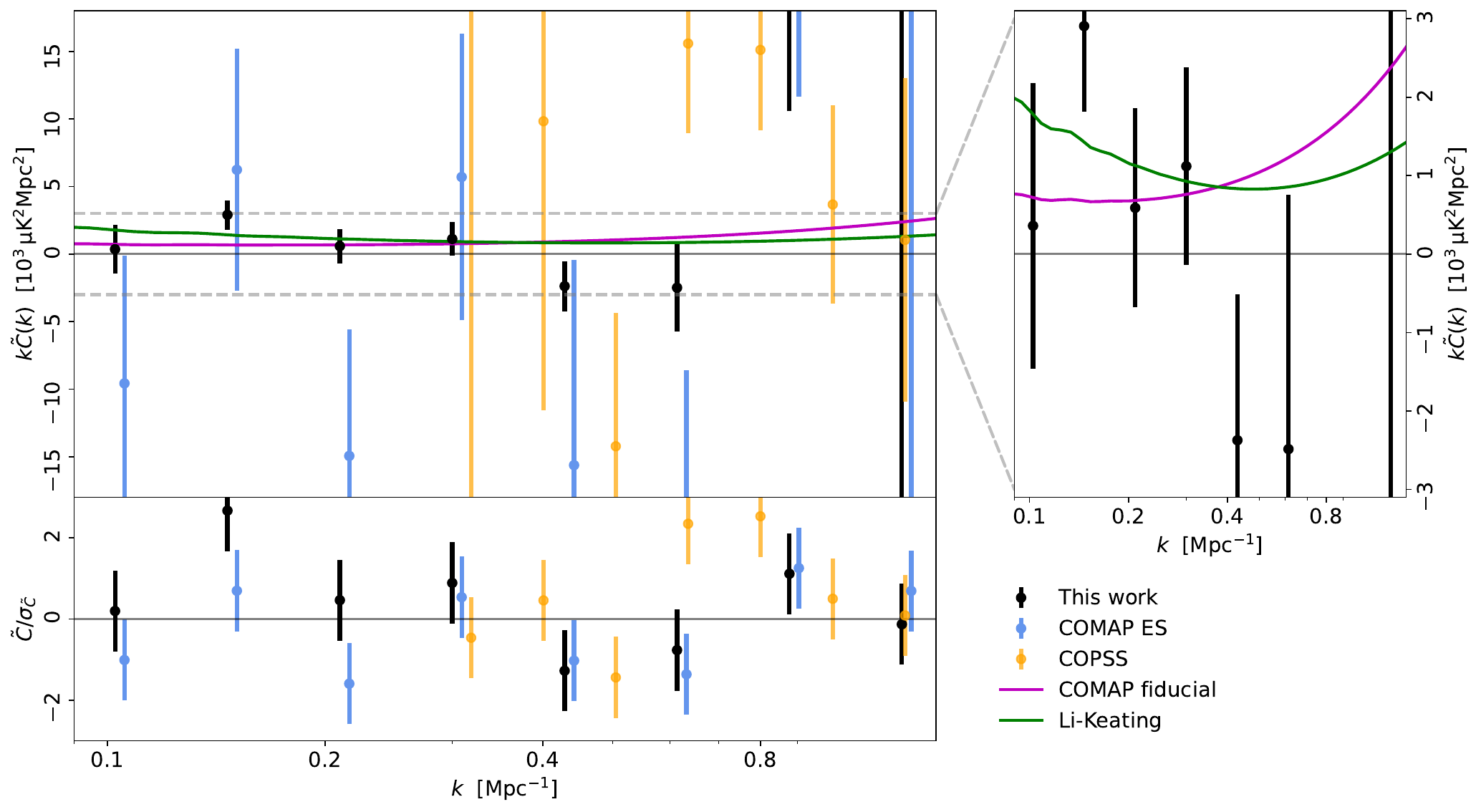}
    \caption{Detailed overview of our field-averaged FGPXS, as well as other datasets and some selected models from the literature. (\textit{Upper panel:}) Spherically averaged FGPXS with $1\sigma$ uncertainties for the field-combined data presented in this paper (black), the COMAP ES field-averaged FPXS (blue; \citealp{comap-IV}), and the COPSS power spectrum (orange; \citealp{keating:2016}). (\textit{Lower panel:}) Corresponding power spectra divided by their respective $1\,\sigma$ uncertainty. (\textit{Inset:}) Zoom-in of the COMAP data points and two comparable models from the literature, namely the fiducial second season COMAP model \citep{comap-V} and the Li-Keating model \cite{li:2016,keating:2020} model. None of the models includes any line-broadening discussed by \cite{chung:2021}. Our data points and those of COMAP ES have been slightly offset from their true $k$-position to increase readability (see Table \ref{tab:power_spectrum values} for an overview over bin centers, FGPXS values and uncertainties).}
    \label{fig:pxs_1d}
\end{figure*}

\subsection{Comparison to COMAP Early Science and COPSS}\label{subsec:comap_ims_vs_es}

Having computed the spherically and field-averaged FGPXS from the data we can compare it to the previous COMAP release as well as the CO Power Spectrum Survey (COPSS), the only other comparable CO(1--0) LIM survey in the literature with published data \citep{keating:2016, kovetz:2017, kovetz:2019, bernal:2022}. This is illustrated in Fig.~\ref{fig:pxs_1d} where the field-averaged FGPXS is plotted together with the COMAP ES constant-elevation-scan FPXS of \cite{comap-IV} and the individual COPSS data points from \cite{keating:2016}. 

The first thing we notice when considering Fig.~\ref{fig:pxs_1d} is the dramatic reduction in the uncertainty of the current measurement compared to that from our ES phase \citep{comap-IV}. Compared to the \cite{comap-IV} FPXS the current level of sensitivity has increased by a factor $\sim 6$--$8$ across our six most sensitive bins at ${0.09\,\mathrm{Mpc}^{-1} < k < 0.73\,\mathrm{Mpc}^{-1}}$. This illustrates the significant increase in the effective data volume by around a factor of eight. Even though the low-level data selection procedure detailed by \cite{lunde:2024} is somewhat more strict than the one in COMAP ES \citep{comap-III, comap-IV}, this is more than compensated by the lack of data cuts in the power spectrum domain, resulting in a significant increase in sensitivity overall. In other words, we have demonstrated that uncertainties in the power spectra integrate down in accordance with expectations for noise-dominated data. 

The two highest $k$-bins have somewhat larger errors in the current result compared to the COMAP ES spectrum. This is due to a combination of the analytical beam transfer function now applied and a stricter 2D $\kvec$-space mask. The beam transfer function now applied is somewhat more strict than the numerically computed one of \cite{comap-IV} on scales closer to the Nyquist limit in the angular direction. Additionally, to avoid problems with aliasing we have masked the outer-most bins in both $k_\parallel$ and $k_\perp$. As a result, the outer-most 1D $k$-bins contain a lower number of samples than they would have for the same 1D bins of \cite{comap-IV}. 

The COPSS power spectrum estimate \citep{keating:2016} primarily covers scales smaller than COMAP, but the two experiments overlap at $0.3\,\mathrm{Mpc}^{-1}\lesssim k \lesssim 1.0\,\mathrm{Mpc}^{-1}$, where they are largely consistent with each other. The only noteworthy disagreement between COPSS and the field-combined FGPXS is a mild $\sim 2.5\sigma$ tension in terms of the combined error between the two power spectrum estimates at $k\sim 0.6\,\mathrm{Mpc}^{-1}$. As we can see from Fig.~\ref{fig:pxs_1d} this point of mild tension coincides with one of the two COPSS points in which they reported a $2.5\sigma$ excess above zero. 

Albeit with large uncertainties, we see that compared to COPSS and COMAP ES the updated COMAP data points cluster significantly closer to, and are consistent with, the two brightest models that were not already excluded in ES \citep{comap-V}, namely the COMAP fiducial model\footnote{A double power-law model relating halo masses in cosmological simulations to CO luminosities; see specifically ``UM+COLDz+COPSS'' in Table 1 of \cite{comap-V} for their fiducial model definition.} and the Li-Keating model of \cite{keating:2020}. For more discussion of the consistency of COPSS with the current COMAP result, including modeling implications, we refer the interested reader to \cite{chung:2024a}.

When comparing the power spectrum sensitivity of COMAP to that of COPSS, we must take into account the smaller $k$-bins in the COPSS analysis. Although the two experiments have a certain region of overlap in $k$-space, the different bin sizes of COPSS and COMAP result in a different intrinsic within-bin variance. To mitigate this effect we can define the normalized sensitivity $\xi_k=\sigma_k \sqrt{\Delta V_k}$, where $\Delta V_k$ is the volume of a spherical $k$-shell defined by the bin $k$ in $k$-space. Two bins with the same value for $\xi_k$ would have the same sensitivity, $\sigma_k$, if they were binned to a standardized bin size. In other words, $\xi_k$ traces the underlying continuous sensitivity of each experiment and $k$-scale, and as a result, the normalized sensitivity across $k$-bins and surveys becomes comparable. We illustrate the normalized sensitivity, $\xi_k$, in Fig.~\ref{fig:norm_sensitivity} for all data points shown in Fig.~\ref{fig:pxs_1d}. The figure nicely illustrates the scales to which COMAP and COPSS are most sensitive. We see that COMAP is most sensitive on large scales at ${0.1\,\mathrm{Mpc}^{-1} < k < 0.3\,\mathrm{Mpc}^{-1}}$, while COPSS is most sensitive on small scales, ${0.5\,\mathrm{Mpc}^{-1} < k < 1.0\,\mathrm{Mpc}^{-1}}$, where the COMAP beam starts to dominate. However, the current FGPXS result has a peak sensitivity increase of around a factor of eight and ten compared to COMAP ES and COPSS respectively (see Table~\ref{tab:normalized_sensitivity} in Appendix~\ref{apdx:norm_sensitivity} for a detailed list of exact normalized sensitivity improvements). This improvement in relative sensitivity compared to COPSS and COMAP ES is expected to increase further as the COMAP instrument gathers more data, and illustrates our ability to remove systematic effects to below the noise level and integrate the noise of the incoming data. In fact, as COPSS to-date remains the only comparable CO(1--0) LIM experiment with published data, COMAP currently provides the most sensitive CO(1--0) LIM constraints in the field.

\subsection{Upper limits on the power spectrum}\label{subsec:ul_results}
Given the factors of, respectively, eight and ten times the sensitivity of our power spectrum result compared to the COMAP ES and COPSS data, it is interesting to consider the upper limits (UL) at $95\,\%$ confidence on a non-zero CO(1--0) power spectrum that can be derived from the data points. These are shown in Fig.~\ref{fig:pxs_1d_ul} for the spherically and field-averaged COMAP FGPXS, the COMAP ES \citep{comap-IV} data points as well as the COPSS \citep{keating:2016} power spectrum estimate. As we are at the level of sensitivity where it becomes more informative to look at the ULs per $k$-bin we only consider the ULs derived per $k$-bin in this work. We show only bin-wise derived ULs from the COMAP ES \citep{comap-IV} and COPSS \citep{keating:2016} data to facilitate a direct comparison to our result. All ULs are computed under the assumption that the astrophysical CO signal must be positive. For comparison, two of the closest models from the literature are included in the plot; the COMAP fiducial model from \cite{comap-V} and the Li-Keating model -- a version of the \cite{li:2016} model from \cite{keating:2020}. While the $0.12\,\mathrm{Mpc}^{-1} < k < 0.18\,\mathrm{Mpc}^{-1}$ FGPXS bin has a $2.7\sigma$ excess above zero, we still present it as a $95\,\%$ upper limit in Fig.~\ref{fig:pxs_1d_ul} as we do not consider the excess statistically significant.   

\begin{figure}[h!]
    \centering
    \includegraphics[width=\linewidth]{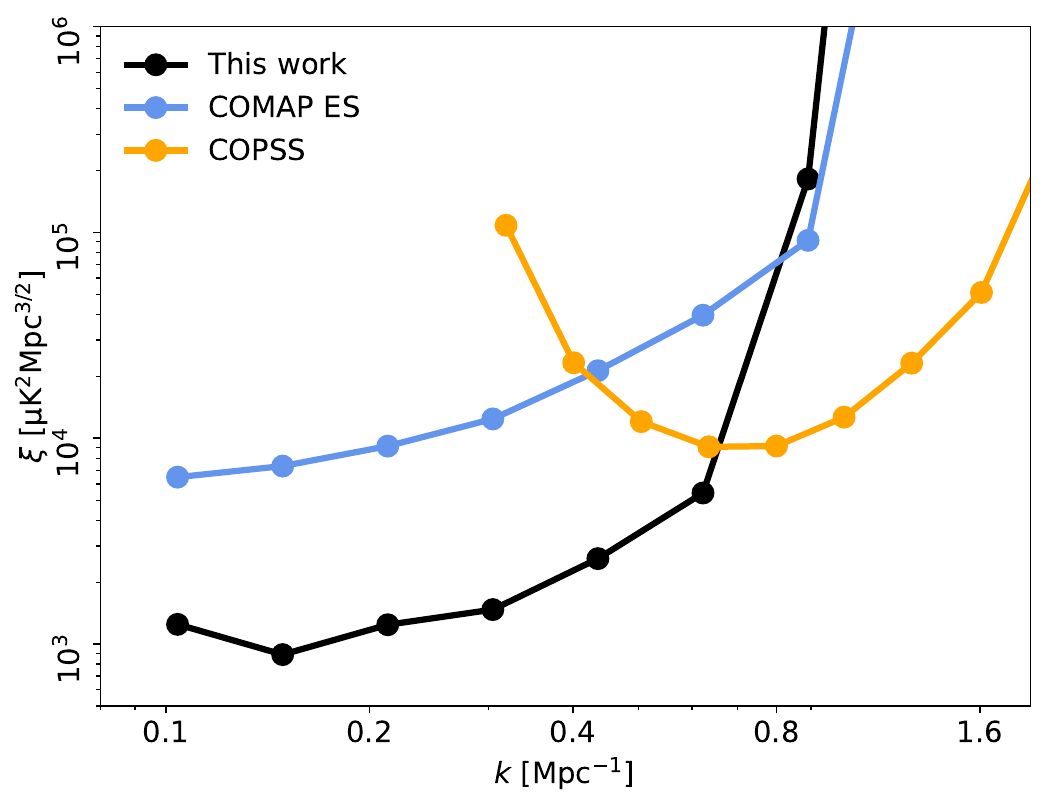}
    \caption{Comparison of volume-normalized sensitivity, $\xi_k$, for the new COMAP FGPXS (black); the previous COMAP ES FPXS (blue; \citealp{comap-IV}), and COPSS (orange; \citealp{keating:2016}). The deepest COMAP $k$-bin is roughly an order of magnitude more sensitive than the deepest COMAP ES and COPSS bins; for tabulated values, see Table~\ref{tab:normalized_sensitivity}.}
    \label{fig:norm_sensitivity}
\end{figure}

\begin{figure*}
    % \centering
    \sidecaption
    \includegraphics[width=12cm]{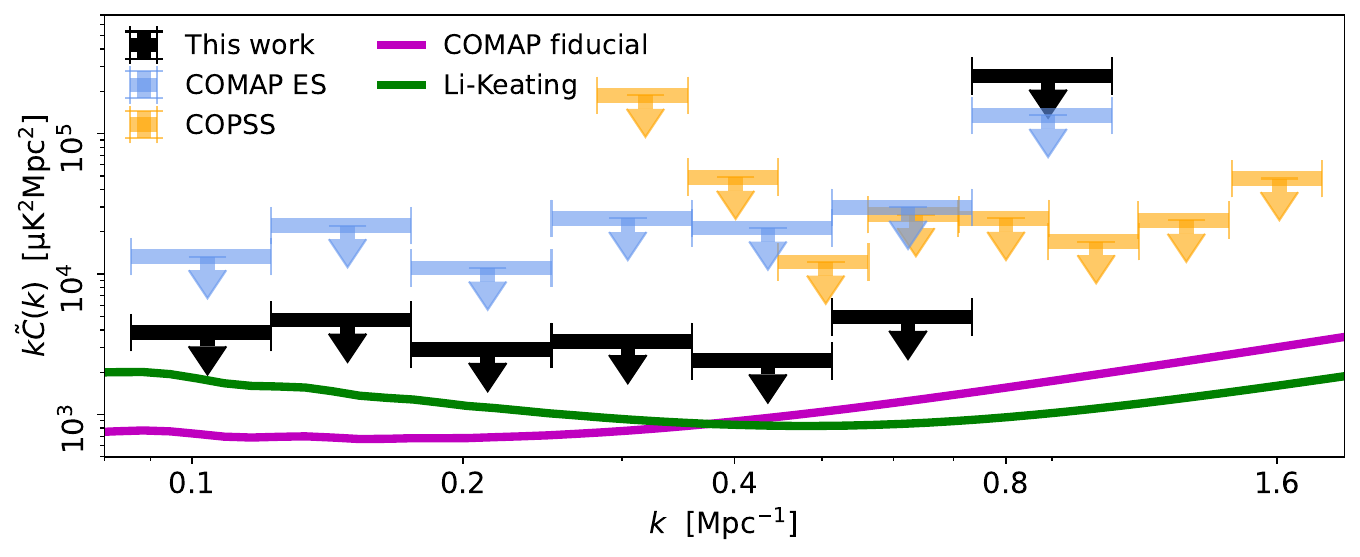}
    \caption{Comparison of upper 95\,\% confidence limits (ULs) on the CO power spectrum as derived from the new COMAP dataset (black), the COMAP ES analysis (blue; \citealp{comap-IV}), and from COPSS (orange; \citealp{keating:2016}). The corresponding data points for each bin are shown in Fig.~\ref{fig:pxs_1d}, and all ULs are derived using a positivity prior. The theoretical model predictions indicated by green and purple lines are the same as in Fig.~\ref{fig:pxs_1d}. We note that because the data point of the FGPXS and COMAP ES centered at $k=1.27\,\mathrm{Mpc}^{-1}$ in Figs.~\ref{fig:pxs_1d_per_field} and \ref{fig:pxs_1d} have large uncertainties the corresponding $95\,\%$ UL are outside $y$-range of the figure.}
    \label{fig:pxs_1d_ul}
\end{figure*}

As in Fig.~\ref{fig:norm_sensitivity}, the ULs we present in Fig.~\ref{fig:pxs_1d_ul} reflect $k$-regions in which each survey is most sensitive. The $95\,\%$ ULs of this work and those derived from COMAP ES \citep{comap-IV} are most constraining in the six most sensitive COMAP bins at $0.09\,\mathrm{Mpc}^{-1} < k < 0.73\,\mathrm{Mpc}^{-1}$. Meanwhile the COPSS \cite{keating:2016} ULs are at their lowest around ${0.7\,\mathrm{Mpc}^{-1} < k < 1.6\,\mathrm{Mpc}^{-1}}$, beyond where the COMAP beam and voxel window dominate and blow up the noise. Compared to COMAP ES, we see a significant improvement in the current ULs per $k$-bin. Specifically, each of our six most sensitive $k$-bins can  individually  constrain ${kP_\mathrm{CO}(k)< 2400-4900\,\mathrm{\mu K^2 Mpc^{2}}}$  at $95\,\%$ confidence. The maximum improvement between the two COMAP releases is around a factor 9 in the $k\sim 0.4\,\mathrm{Mpc}^{-1}$ bin. The UL estimates are sensitive to both the uncertainty of a data point and its value. As the field-averaged FGPXS in the $k\sim 0.4\,\mathrm{Mpc}^{-1}$ bin is around $-1.5\sigma$ below zero the resulting UL becomes the deepest even though according to Fig.~\ref{fig:norm_sensitivity} it is not the most sensitive $k$-bin.

When comparing COPSS to COMAP in Fig.~\ref{fig:pxs_1d_ul} we see that where COMAP and COPSS have overlapping areas of high sensitivity, at $k<0.8\,\mathrm{Mpc}^{-1}$, our $95\,\%$ ULs are significantly lower than those derived from the COPSS data points. This reflects the increased sensitivity of the COMAP FGPXS estimate already observed in Fig.~\ref{fig:norm_sensitivity}.  While none of the updated $95\,\%$ ULs are touching any of the two included models, a significantly larger region of the power spectrum space is excluded compared to only using the COPSS and COMAP ES limits, and our $95\,\%$ ULs are starting to encroach on the models that are not already excluded, including the fiducial model \citep{comap-V}. Given our demonstrated ability to control systematic effects, and the constraints already achieved, detection of a CO power spectrum close to the fiducial model is within reach with further observations.

To conclude the discussion of the power spectrum results, the current COMAP power spectrum is the state-of-the-art CO LIM power spectrum dataset with around an order of magnitude more sensitivity and comparatively lower ULs at $95\,\%$ confidence than COPSS and COMAP ES, the only comparable CO(1--0) line intensity mapping datasets in the literature. The presented power spectrum data points and resulting $95\,\%$ ULs further exclude a significant portion of the parameter space of possible CO models and provide the current best direct 3D constraints on the CO(1--0) power spectrum in the literature \citep{kovetz:2017, kovetz:2019, bernal:2022}.

\section{Null test results}\label{sec:null_test_results}
As was described in Sect.~\ref{sec:null_tests_methodology}, we performed a set of null tests by computing the average cross-elevation FGPXS of a set of difference maps. All null tests were performed with the same pipeline and data selection as the power spectrum data shown in Sect.~\ref{sec:power_spectrum_results}.  The differencing variables chosen for the null tests were selected to test for correlations owing to a variety of potential systematic effects, for example, environmental effects like weather, sidelobe pickup, and pipeline diagnostics. In Table~\ref{tab:null_variables} we show an overview of the selected null variables.

In total, 312 effective null tests were performed: 26 null test variables across three fields, cylindrical- and spherical-averaged FGPXS as well as separate tests for fast and slow azimuth data respectively. All of these can have different associated systematic effects. For instance, given that the telescope's scanning speed was changed to a lower azimuthal speed in May 2022, the fast and slow azimuth data (May 2022 -- November 2023) may have very different mechanical vibrations that could cause spurious patterns in the maps \citep[see][for examples]{lunde:2024}.

For each of the effective null tests, we calculate corresponding $\chi^2$ PTEs, as is described in Sect.~\ref{sec:null_tests_methodology}. We provide a detailed list of these in Appendix~\ref{apdx:null_results} (see Table~\ref{tab:pte}). 

Of the 312 null tests that we performed, the two lowest PTEs were found to be $\sim 0.6\,\%$, which amounts to a random binomial probability of $27\,\%$. Two of the null test $\chi^2$-values were slightly outside the RND simulated $\chi^2$-distribution and we therefore only have a lower limit of $99.5\,\%$ on their PTEs (because the numerical resolution of the simulation-based approach is $1/183\approx0.5\,\%$); this could be improved somewhat by using more RND realizations. 

\begin{figure}
    \centering
    \includegraphics[width=\linewidth]{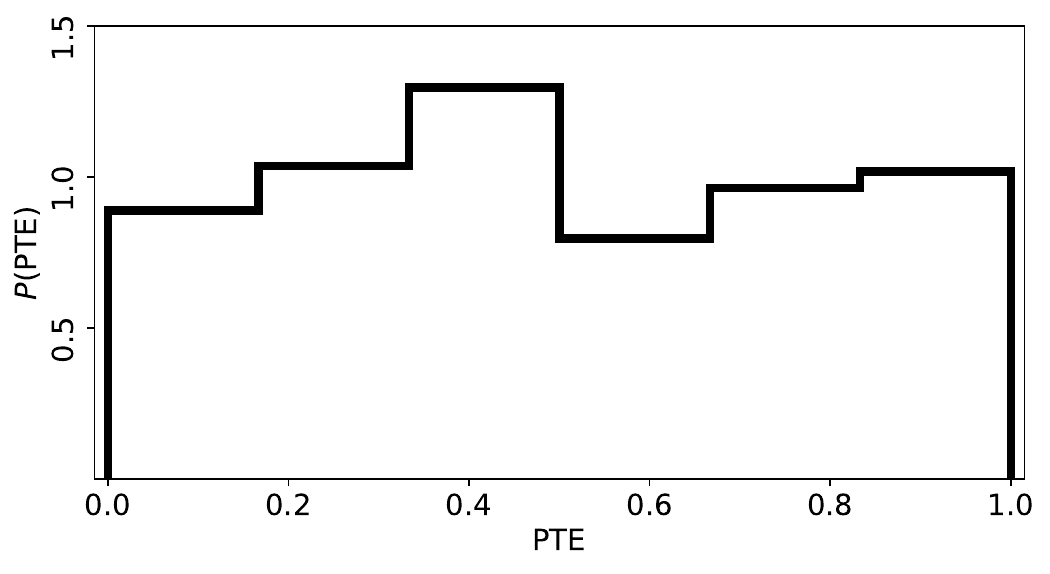}
    \caption{Normalized distribution, $P(\mathrm{PTE})$, of $\chi^2$ PTEs for all null tests performed on  Field 1-3 combined. The PTE values corresponding to this histogram are found in Table~\ref{tab:pte} in Appendix~\ref{apdx:null_results}. The Kolmogorov-Smirnov (KS) uniformity test on the samples contained in the illustrated distribution was found to yield a KS PTE of $58.7\,\%$ (see Table~\ref{tab:ks}).}
    \label{fig:pte_histogram_combined}
\end{figure}

The PTEs are expected to follow a uniform distribution. As a consistency check, we therefore consider the PTE distributions of the performed null tests. In Fig.~\ref{fig:pte_histogram_combined} we show the combined PTE distribution for all separately performed null tests (the corresponding distributions for each separately performed category of null tests can be seen in Fig.~\ref{fig:pte_histograms} of Appendix~\ref{apdx:null_results}). To further gauge the uniformity of the histograms a Kolmogorov-Smirnov (KS) test was performed to see if the null test $\chi^2$ PTEs were consistent with the null hypothesis of being drawn from a uniform distribution. The KS-test PTE-values are found in Table~\ref{tab:ks} (and also in the bottom row of Table~\ref{tab:pte} of null test $\chi^2$ PTEs in Appendix~\ref{apdx:null_results}). The lowest KS-test PTE of $5.5\,\%$, corresponds to a binomial probability of around  $35\,\%$ for the 12 performed KS tests. The maximum KS-test PTE is around $79\,\%$, and the uniformity of the entire set of PTEs is at the $58.7\,\%$ level. 

We can therefore conclude that all the null tests and PTE uniformity tests have been passed and are consistent with the expected instrumental noise. As we do not claim any detection at this stage, the number and type of null tests performed are more than enough to ensure a sufficient data quality for our upper limits. 

\begin{table*}
    \caption{Kolmogorov-Smirnov uniformity test PTEs on the null test $\chi^2$ PTEs.}
    \begin{center}
    \begin{tabular}{L{2cm}L{0.8cm}L{0.8cm}L{0.8cm}L{0.8cm}L{0.8cm}L{1.1cm}L{1.1cm}L{0.8cm}L{0.8cm}L{0.8cm}L{0.8cm}L{0.8cm}}
    \hline \hline 
    & \multicolumn{12}{c}{{Kolmogorov-Smirnov probabilities-to-exceed (KS PTEs)} [\%]}\\
    \hline
    & \multicolumn{6}{c}{{Spherically averaged (1D)}} & \multicolumn{6}{c}{{Cylindrically averaged (2D)}}\\
    \hline
    & \multicolumn{2}{c}{{Field 1}} 
    & \multicolumn{2}{c}{{Field 2}} 
    & \multicolumn{2}{c}{{Field 3}} 
    & \multicolumn{2}{c}{{Field 1}} 
    & \multicolumn{2}{c}{{Field 2}}
    & \multicolumn{2}{c}{{Field 3}}\\
    \hline
    {Combined} & {Fast} & {Slow} & {Fast} & {Slow} & {Fast} & {Slow} & {Fast} & {Slow} & {Fast} & {Slow} & {Fast} & {Slow} \\
    \hline
    58.7 & 5.5 & 9.7 & 16.9 & 24.1 & 41.8 & 48.9 & 32.1 & 8.4 & 61.9 & 78.7 & 70.9 & 72.0 \\
    \hline
    \end{tabular}
    \tablefoot{Probabilities-to-exceed of Kolmogorov-Smirnov (KS PTEs) uniformity test of the null tests $\chi^2$ PTEs (found in Table~\ref{tab:pte} of Appendix~\ref{apdx:null_results}) in units percent of all three Fields, fast- and slow-moving azimuth scans (denoted as ``Fast'' and ``Slow'') as well as the spherically and cylindrically averaged FGPXS.}
    \label{tab:ks}
    \end{center}
\end{table*}

\section{Conclusion}\label{sec:conclusion}

We have presented updated constraints on the cosmological CO(1--0) power spectrum at $2.4 < z < 3.4$, derived from the latest COMAP observations. These measurements are based on a novel mean-averaged feed group pseudo-cross-power spectrum (FGPXS) estimator that is a slight modification of the feed-feed pseudo-cross-power spectrum (FPXS) estimator used in the COMAP ES analysis \citep{comap-IV}. The difference between these two estimators is that while the previous estimator evaluated cross-correlations between any two detector feeds, the new estimator evaluates cross-correlations between groups of feeds defined by common first downconversion (DCM1) local oscillators. The motivation for this is that feeds in these groups share some common instrumental systematic effects, and the new estimator is therefore more robust against such effects.

Quantitatively, all power spectrum bins were consistent with zero up to $\sim 2\sigma$, except for $k\sim 0.15\,\mathrm{Mpc}^{-1}$, which showed a $2$--$3\sigma$ excess in Fields 1 and 2; averaging over all three fields yields an excess of $2.7\sigma$. Despite this single-bin excess, the total PTEs with respect to a zero-signal model are  $33.2\,\%$, $19.5\,\%$ and $82.7\,\%$ for Fields 1--3, respectively, and $34\,\%$ when combining the data across fields. The resulting FGPXS spectrum derived from the latest COMAP data is thus statistically consistent with instrumental noise, and a detailed suite of null tests shows no signs of residual systematic effects. At the same time, the slight excess at $k\sim 0.15\,\mathrm{Mpc}^{-1}$ is noteworthy; it could just be a regular noise fluctuation or the signature of some yet-to-be-discovered systematic effect. However, it could also be a small first hint of true cosmological CO fluctuations. More data are needed to determine its true nature.

Comparing with previous results, we find that the new COMAP power constraints are almost an order of magnitude stronger than the previous ES results \citep{comap-IV}. In addition, when considering the power spectrum data points alone, the COPSS power spectrum \citep{keating:2016} was found to be mostly consistent with the COMAP FGPXS, with only a mild $\sim 2.5\,\sigma$ tension in one of the bins. The volume-normalized sensitivity of the COMAP FGPXS was found to be around ten times that of the COPSS power spectrum estimate when comparing the respective most sensitive bins of the two experiments. 

We developed a null test framework involving the difference between half-data maps that are split under variables believed to be associated with systematic effects. With the 26 split variables, three fields, the cylindrically and spherically averaged FGPXS as well as the fast- and slow-moving scans a total of 312 effective null tests were performed. Of these all passed within the expected instrumental uncertainties, ensuring the quality of our final data products.

To conclude, our power spectrum estimates and the resulting $95\,\%$ upper limits provide the most sensitive constraints on cosmic CO emission at $z\sim 2-3$ published to date and significantly reduce the allowed parameter space of possible CO emission models, the implications of which we explore further in the companion work of~\cite{chung:2024a}. These results are a strong demonstration of COMAP's powerful capabilities and performance in terms of systematic effect mitigation, and the filtered data are still dominated by white noise even after three years of integration. Regular operations are still ongoing, and the data currently being gathered will put further pressure on possible CO emission models. 

\begin{acknowledgements}    
    We acknowledge support from the Research Council of Norway through
    grants 251328 and 274990, and from the European Research Council (ERC)
    under the Horizon 2020 Research and Innovation Program (Grant agreement
    No. 772253 bits2cosmology and 819478 Cosmoglobe).
    
    This material is based upon work supported by the National Science Foundation under Grant Nos.\ 1517108, 1517288, 1517598, 1518282, 1910999, and 2206834, as well as by the Keck Institute for Space Studies under ``The First Billion Years: A Technical Development Program for Spectral Line Observations''.

    Parts of the work were carried out at the Jet Propulsion Laboratory, California Institute of Technology, under a contract with the National Aeronautics and Space Administration.

    PCB is supported by the James Arthur Postdoctoral Fellowship

    HP acknowledges support from the Swiss National Science Foundation via Ambizione Grant PZ00P2\_179934. 

    JK acknowledges support from a Robert A. Millikan Fellowship while at Caltech.

    DC is supported by a CITA/Dunlap Institute postdoctoral fellowship. The Dunlap Institute is funded through an endowment established by the David Dunlap family and the University of Toronto. Research in Canada is supported by NSERC and CIFAR.

    SEH acknowledges funding from an STFC Consolidated Grant (ST/P000649/1) and a UKSA grant (ST/Y005945/1) funding LiteBIRD foreground activities.

    JG acknowledges support from the Keck Institute for Space Science, NSF AST-1517108, and University of Miami, and Hugh Medrano for assistance with cryostat design.

    This work was supported by the National Research Foundation of Korea(NRF) grant funded by the Korean government(MSIT) (RS-2024-00340759).

    NOS and JGSL thank Sigurd~K.~N{\ae}ss for all the fruitful discussions in the office, and while biking through nature, during the last three years.

    This work was first presented at the Line Intensity Mapping 2024 conference held in Urbana, Illinois; we thank the organizers for their hospitality and the participants for useful discussions.
    
    \newline\newline \textit{Software acknowledgments}. Matplotlib for plotting  \citep{matplotlib:2007}; NumPy \citep{numpy:2020} and SciPy \citep{scipy:2020} for efficient numerics and array handling in Python; Astropy a community-made core Python package for astronomy \citep{astropy:2013, astropy:2018, astropy:2022}; Multi-node parallelization with MPI for Python \citep{mpi4py:2005, mpi4py:2008, mpi4py:2011, mpi4py:2021}; Pixell (\url{https://github.com/simonsobs/pixell}) for handling sky maps in rectangular pixelization.
\end{acknowledgements}

\bibliographystyle{aa}
\bibliography{references}

\begin{appendix}

\section{Normalized CO power spectrum sensitivity}\label{apdx:norm_sensitivity}
Table~\ref{tab:normalized_sensitivity} provides a comparison of the volume-normalized CO power spectrum sensitivity for the new COMAP dataset with those power spectra derived from COMAP ES and COPSS; these are visualized in Fig.~\ref{fig:norm_sensitivity}. The volume-normalized sensitivity is defined as  
\begin{equation}
    \xi_k=\sigma_k \sqrt{\Delta V_k},
\end{equation}
where $\sigma_k$ is the uncertainty of a spherically averaged power spectrum bin $k$ with shell volume $\Delta V_k$. This definition eliminates the effect of within-bin variance at each $k$-bin and provides a volume-independent measure that may be used to compare sensitivities between non-overlapping power spectrum bins and surveys. We see that the current COMAP power spectrum constraints reach a maximum sensitivity of one order of magnitude higher than the most sensitive COPSS \citep{keating:2016} and COMAP ES \citep{comap-IV} bins. In addition, it is important to note that the regimes of maximum sensitivity differ between COMAP and COPSS, and this is due to their different instrumental designs and effective angular resolutions; COMAP is more sensitive in the large-scale clustering regime, while COPSS is more sensitive in the small-scale shot-noise regime. 
\begin{table*}
    \caption{\label{tab:normalized_sensitivity} Normalized sensitivity in each COMAP and COPSS bin.}
    \begin{center}
        \begin{tabular}{Z{2cm}Z{1.5cm}Z{1.5cm}Z{1.5cm}S{3.5cm}S{2cm}}
        \hline \hline 
        \textbf{Survey} & \textbf{$k$-center [$\mathrm{Mpc}^{-1}$]} & \textbf{$k$-min [$\mathrm{Mpc}^{-1}$]} & \textbf{$k$-max [$\mathrm{Mpc}^{-1}$]} & \textbf{normalized sensitivity $\xi_k$ [$10^3\mathrm{\mu K}^{2}\mathrm{Mpc}^{3/2}$]} & \textbf{$\xi_\mathrm{COPSS}^\mathrm{min} / \xi_k$} \\ 
        \hline
        \multirow{7}{2cm}{\textbf{This work}}
        & 0.1 & 0.09 & 0.12 & 1.25 & 7.3 \\
        & 0.15 & 0.12 & 0.18 & 0.89 & 10.2 \\
        & 0.21 & 0.18 & 0.25 & 1.24 & 7.3 \\
        & 0.3 & 0.25 & 0.36 & 1.47 & 6.2 \\
        & 0.44 & 0.36 & 0.51 & 2.6 & 3.5 \\
        & 0.62 & 0.51 & 0.73 & 5.42 & 1.7 \\
        & 0.89 & 0.73 & 1.05 & 181.92 & 0.0499 \\
        & 1.27 & 1.05 & 1.5 & $9.3\times 10^6$ & $9.7\times 10^{-7}$ \\
        \hline
        \multirow{7}{2cm}{\textbf{COMAP ES}}
        & 0.1 & 0.09 & 0.12 & 6.47 & 1.4 \\
        & 0.15 & 0.12 & 0.18 & 7.32 & 1.2 \\
        & 0.21 & 0.18 & 0.25 & 9.16 & 1.0 \\
        & 0.3 & 0.25 & 0.36 & 12.41 & 0.7 \\
        & 0.44 & 0.36 & 0.51 & 21.28 & 0.4 \\
        & 0.62 & 0.51 & 0.73 & 39.6 & 0.2 \\
        & 0.89 & 0.73 & 1.05 & 91.5 & 0.0993 \\
        & 1.27 & 1.05 & 1.5 & 28000 & 0.0003 \\
        \hline
        \multirow{7}{2cm}{\textbf{COPSS bins}}
        & 0.4 & 0.36 & 0.45 & 23.27 & 0.39 \\
        & 0.5 & 0.45 & 0.57 & 12.04 & 0.754 \\
        & 0.64 & 0.57 & 0.71 & 9.08 & 1.0 \\
        & 0.8 & 0.71 & 0.9 & 9.16 & 0.991 \\
        & 1.01 & 0.9 & 1.13 & 12.66 & 0.718 \\
        & 1.27 & 1.13 & 1.42 & 23.16 & 0.392 \\
        & 1.61 & 1.42 & 1.79 & 51.08 & 0.178 \\
        & 2.02 & 1.79 & 2.84 & 277.81 & 0.033 \\
        & 3.2 & 2.84 & 3.57 & 2060.78 & 0.004 \\
        \hline
        \end{tabular}
    \tablefoot{Volume-normalized sensitivity, $\xi_k$, of each of our field-averaged power spectrum bins as well as the COPSS measurement \citep{keating:2016}. The normalized sensitivity ratio of COMAP (i.e., this work), COMAP ES \citep{comap-IV} and the individual COPSS bins relative to the most sensitive COPSS bin \citep{keating:2016} is given by $\xi_\mathrm{COPSS}^\mathrm{min}/\xi_k$.}
    \end{center}
\end{table*}

\FloatBarrier
\section{Power spectrum data point values}\label{apdx:psx_points}
For the interested reader, we provide a list of power spectrum values and uncertainties, $k\tilde{C}(k)$ and $k\sigma_{\tilde{C}(k)}$ respectively, of the spherically and field-averaged FGPXS data points seen in Fig.~\ref{fig:pxs_1d}. These can be found in Table~\ref{tab:power_spectrum values}.

\begin{table}[h!]
    \caption{\label{tab:power_spectrum values} Overview of FGPXS bin values and uncertainties.}
    \begin{center}
        \begin{tabular}{Z{2.3cm}S{2.3cm}S{2.3cm}}
        \hline \hline 
        $k$-center [$\mathrm{Mpc}^{-1}$] & $k\tilde{C}(k)$ [$10^3\mathrm{\mu K}^2\mathrm{Mpc}^{2}$] & $k\sigma_{\tilde{C}(k)}$ [$10^3\mathrm{\mu K}^2\mathrm{Mpc}^{2}$] \\
        \hline
        $0.1$\dotfill &  $0.36$ & $1.82$ \\
        $0.15$\dotfill &  $2.9$ & $1.09$\\
        $0.21$\dotfill & $0.59$ & $1.27$\\
        $0.3$\dotfill &  $1.19$ & $1.26$\\
        $0.44$\dotfill &  $-2.37$ & $1.86$\\
        $0.62$\dotfill &  $-2.48$ & $3.24$\\
        $0.89$\dotfill &  $101.5$ & $90.9$\\ 
        $1.27$\dotfill & $-5.05\times 10^5$ & $3.9\times 10^6$\\
        \hline
        \end{tabular}
    \tablefoot{Bin values and uncertainties (respectively in the last two columns) of the spherically  and field-averaged FGPXS corresponding to our data points seen in Fig. \ref{fig:pxs_1d}.}
    \end{center}
\end{table}

\FloatBarrier
\section{Null test probabilities-to-exceed}\label{apdx:null_results}
In the following, we present a summary of the $\chi^2$ PTEs for each of our effective 312 null tests performed. The PTEs are found in Table~\ref{tab:pte} and each null variable, and the corresponding acronyms are explained in detail in Table~\ref{tab:null_variables}. 

Table~\ref{tab:pte} is structured as follows: each row shows a different null variable in which the data was split in two, such as ambient temperature (ambt) or right- and left-moving azimuth sweeps (azdr). The columns are grouped into a hierarchical structure, as we performed null tests separately on spherically and cylindrically averaged FGPXS, for each field (Fields 1-3) as well as for data that were gathered before and after May 2022 when the scanning speed of the telescope was reduced. That is, because the fast- and slow-moving azimuth scans may have different associated systematic effects from, for example, mechanical vibrations in the telescope. 

\begin{figure}
    \centering
    \includegraphics[width=\linewidth]{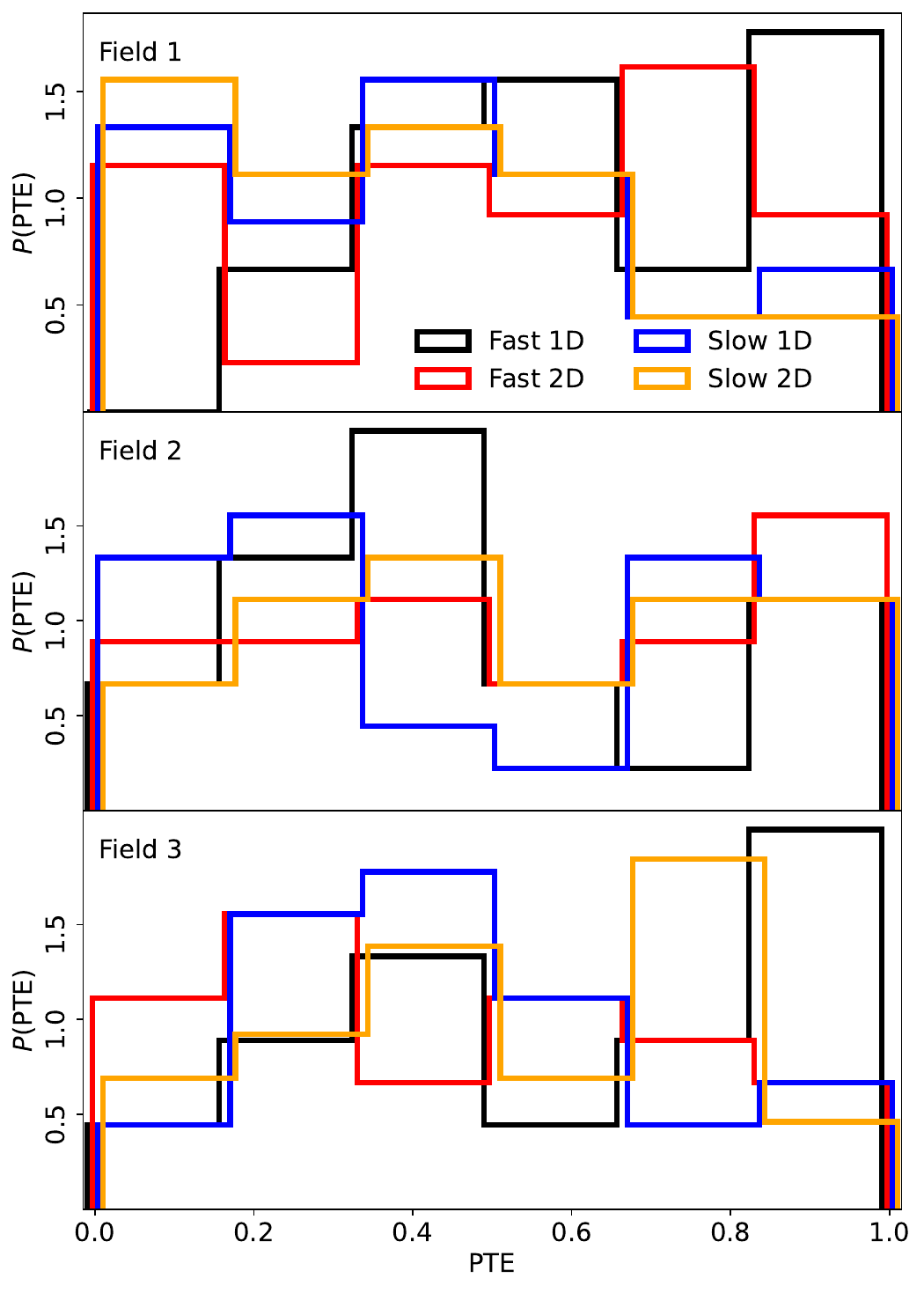}
    \caption{Normalized distribution, $P(\mathrm{PTE})$, of $\chi^2$ PTEs separately for all null tests performed on  Field 1-3. The PTE values corresponding to this histogram are found in Table~\ref{tab:pte}. The Kolmogorov-Smirnov (KS) uniformity test results for the $\chi^2$-samples of each of the separate PTE distributions can be found in Table~\ref{tab:ks} or the bottom row of Table~\ref{tab:pte}. The individual histograms are slightly offset w.r.t. each other for increased readability.}
    \label{fig:pte_histograms}
\end{figure}

We present the distributions of PTEs of each separately performed null test in Fig.~\ref{fig:pte_histograms} (see Fig.~\ref{fig:pte_histogram_combined} in Sect.~\ref{sec:null_test_results} for distribution of all null test PTEs considered jointly). As the distribution of PTEs is expected to be uniform we also performed a Kolmogorov-Smirnov (KS) test to find how probable it is that the PTE samples are drawn from a uniform distribution. These are shown for each separate null test category in the very last row of Table~\ref{tab:pte} (and also in Table~\ref{tab:ks} in Sect.~\ref{sec:null_test_results}). For a discussion on the null test results see Sect.~\ref{sec:null_test_results}.

\begin{table*}
    \caption{\label{tab:pte} Detailed overview of null test $\chi^2$ PTEs.}
    \begin{center}
    \begin{tabular}{L{2cm}L{0.8cm}L{0.8cm}L{0.8cm}L{0.8cm}L{0.8cm}L{1.2cm}L{1.2cm}L{0.8cm}L{0.8cm}L{0.8cm}L{0.8cm}L{0.8cm}}
    \hline \hline 
    & \multicolumn{12}{c}{\textbf{$\chi^2$ probabilities-to-exceed} [\%]}\\
    \hline
    & \multicolumn{6}{c}{\textbf{Spherically averaged (1D)}} & \multicolumn{6}{c}{\textbf{Cylindrically averaged (2D)}}\\
    \hline
    & \multicolumn{2}{c}{\textbf{Field 1}} 
    & \multicolumn{2}{c}{\textbf{Field 2}} 
    & \multicolumn{2}{c}{\textbf{Field 3}} 
    & \multicolumn{2}{c}{\textbf{Field 1}} 
    & \multicolumn{2}{c}{\textbf{Field 2}}
    & \multicolumn{2}{c}{\textbf{Field 3}}\\
    \hline
    \textbf{Null variable} & \textbf{Fast} & \textbf{Slow} & \textbf{Fast} & \textbf{Slow} & \textbf{Fast} & \textbf{Slow} & \textbf{Fast} & \textbf{Slow} & \textbf{Fast} & \textbf{Slow} & \textbf{Fast} & \textbf{Slow} \\
    \hline
    ambt\dotfill & 57 & 33 & 8 & 13 & 76 & 22 & 77 & 45 & 49 & 26 & 70 & 74 \\
    wind\dotfill & 43 & 17 & 46 & 29 & 83 & 40 & 77 & 7 & 45 & 26 & 17 & 87 \\
    wint\dotfill & 41 & 49 & 64 & 93 & 17 & 62 & 7 & 24 & 90 & 91 & 47 & 76 \\
    half\dotfill & 86 & 16 & 21 & 6 & 14 & 43 & 49 & 3 & 87 & 39 & 25 & 68 \\
    odde\dotfill & 86 & 48 & 38 & 91 & 38 & 66 & 54 & 14 & 9 & 91 & 95 & 43 \\
    dayn\dotfill & 61 & 87 & 86 & 77 & 29 & 11 & 73 & 7 & 76 & 79 & 3 & 79 \\
    dtmp\dotfill & 25 & 19 & 38 & 49 & 92 & $\geq 99.5\tablefootmark{a}$ & 14 & 27 & 36 & 83 & 27 & 42 \\
    hmty\dotfill & 95 & 52 & 98 & 70 & 51 & 17 & 90 & 41 & 81 & 79 & 29 & 44 \\
    pres\dotfill & 52 & 92 & 32 & 23 & 35 & 76 & 9 & 43 & 2 & 99 & 61 & 20 \\
    wthr\dotfill & 69 & 79 & 67 & 37 & 38 & 28 & 37 & 15 & 45 & 62 & 63 & 68 \\
    sune\dotfill & 89 & 58 & 95 & 27 & 96 & 28 & 40 & 78 & 79 & 8 & 7 & 42 \\
    modi\dotfill & 34 & 46 & 16 & 27 & 72 & 48 & 83 & 51 & 88 & 42 & 90 & 37 \\
    sudi\dotfill & 91 & 55 & 16 & 27 & 97 & 94 & 26 & 24 & 20 & 96 & 44 & 80 \\
    tsys\dotfill & 26 & 42 & 33 & 93 & 17 & 28 & 9 & 88 & 61 & 44 & 54 & 93 \\
    fpoO\dotfill & 44 & 97 & 11 & 3 & 64 & 62 & 60 & 56 & 62 & 83 & 74 & 52 \\
    fpoI\dotfill & 39 & 45 & 44 & 73 & 49 & 3 & 62 & 74 & 44 & 89 & 53 & 25 \\
    apoO\dotfill & 50 & 72 & 42 & 92 & 86 & 44 & 60 & 98 & 73 & 27 & 17 & 0.6 \\
    apoI\dotfill & 68 & 58 & 87 & 71 & 67 & 25 & $\geq 99.5\tablefootmark{a}$ & 18 & 1 & 8 & 76 & 21 \\
    spoO\dotfill & 74 & 17 & 95 & 76 & 24 & 61 & 81 & 38 & 92 & 57 & 97 & 69 \\
    spoI\dotfill & 51 & 9 & 37 & 27 & 70 & 37 & 96 & 58 & 57 & 13 & 11 & 11 \\
    npca\dotfill & 89 & 51 & 19 & 80 & 86 & 98 & 92 & 49 & 16 & 45 & 74 & 55 \\
    pcaa\dotfill & 58 & 30 & 64 & 60 & 97 & 19 & 35 & 67 & 27 & 42 & 32 & 24 \\
    s01f\dotfill & 61 & 42 & 33 & 16 & 44 & 37 & 88 & 3 & 89 & 24 & 60 & 80 \\
    fk1f\dotfill & 85 & 16 & 1 & 20 & 36 & 49 & 78 & 56 & 1 & 68 & 11 & 17 \\
    al1f\dotfill & 30 & 14 & 50 & 16 & 83 & 62 & 81 & 48 & 90 & 67 & 49 & 39 \\
    azdr\dotfill & 43 & 42 & 34 & 98 & 93 & 71 & 0.6 & 27 & 18 & 20 & 11 & 3 \\
    \hline
    \textbf{KS-test}\dotfill & 5.5 & 9.7 & 16.9 & 24.1 & 41.8 & 48.9 & 32.1 & 8.4 & 61.9 & 78.7 & 70.9 & 72.0 \\
    \hline
    \end{tabular}
    \tablefoot{Null test $\chi^2$ PTEs in units percent. All tabulated PTE values, for all three Fields, fast- and slow-moving azimuth scans (denoted as ``Fast'' and ``Slow''), are numerically computed from the RND ensemble. The last row indicates the Kolmogorov-Smirnov (KS) uniformity test PTE. The KS uniformity PTE of the entire table of PTEs is $58.7\,\%$. \\
    \tablefoottext{a}{ The $\chi^2$-value of this null test was slightly outside the simulated RND $\chi^2$-distribution and we hence only have a lower limit of $99.5\,\%$ on the numerical PTE as the numerical resolution of the simulated distribution is roughly $1/183 \sim 0.5\,\%$ from the RND ensemble size. }
    }
    \end{center}
\end{table*}

\begin{table*}
    \caption{\label{tab:null_variables} Detailed overview and explanation of null test variables.}
    \begin{tabular}{L{4cm}l}
    \hline \hline 
    \textbf{Null variable} & \textbf{Explanation} \\ 
    \hline
    ambt\dotfill & Ambient temperature at telescope site, as recorded by a nearby weather station.\\
    wind\dotfill & Wind speed, as recorded by a nearby weather station. \\
    wint\dotfill & Winter/summer split, the time difference to the middle of winter (15th of January).\\
    half\dotfill & Half-mission split, early versus late scans.\\
    odde\dotfill & Odd versus even scans.\\
    dayn\dotfill & Day/night split, time difference to 2 AM.\\
    dtmp\dotfill & Dew temperature, as recorded by a nearby weather station.\\
    hmty\dotfill & Humidity, as recorded by a nearby weather station.\\
    pres\dotfill & Air pressure, as recorded by a nearby weather station.\\
    wthr\dotfill & Bad weather and cloud coverage, predicted by a neural network trained on the raw data. \\
    sune\dotfill & Sun elevation.\\
    modi\dotfill & Average angular distance from the center of the field to the moon during the scan.\\
    sudi\dotfill & Average angular distance from the center of the field to the sun during the scan.\\
    tsys\dotfill & Average system temperature, as measured by the vane calibration, during the scan.\\
    fpo0\dotfill & $f_\mathrm{knee}$ value of a $1/f$ fit on the 0th-order $1/f$ gain fluctuation filter coefficient.\tablefootmark{a}\\
    fpo1\dotfill & $f_\mathrm{knee}$ value of a $1/f$ fit on the 1st-order $1/f$ gain fluctuation filter coefficient.\tablefootmark{a}\\
    apo0\dotfill & $\alpha$ value of a $1/f$ fit on the 0th-order $1/f$ gain fluctuation filter coefficient.\tablefootmark{a}\\
    apo1\dotfill & $\alpha$ value of a $1/f$ fit on the 1st-order $1/f$ gain fluctuation filter coefficient.\tablefootmark{a}\\
    spo0\dotfill & $\sigma_0$ value of a $1/f$ fit on the 0th-order $1/f$ gain fluctuation filter coefficient.\tablefootmark{a}\\
    spo1\dotfill & $\sigma_0$ value of a $1/f$ fit on the 1th-order $1/f$ gain fluctuation filter coefficient.\tablefootmark{a}\\
    npca\dotfill & Number of PCA components subtracted in the TOD filtering pipeline.\\
    pcaa\dotfill & Average amplitude of the fitted PCA components in the TOD filtering pipeline.\\
    s01f\dotfill & $\sigma_0$ value of a $1/f$ fit on the sideband-averaged time-domain data.\\
    fk1f\dotfill & $f_\mathrm{knee}$ value of a $1/f$ fit on the sideband-averaged time-domain data.\\
    al1f\dotfill & $\alpha$ value of a $1/f$ fit on the sideband-averaged time-domain data.\\
    azdr\dotfill & Scans split internally in left- vs right-moving pointing, in azimuth.\\
    \hline
    \end{tabular}
    \tablefoot{Explanation of the null test split variables. For all variables, we show the abbreviation used in Table \ref{tab:pte} and a more detailed explanation of the null test variable\\
    \tablefoottext{a}{As part of the TOD filtering a first order polynomial is fitted across the frequency bands for each time-sample. The 0th- and 1st-order polynomial components (as functions of time) tend to follow a $1/f$ spectrum, and a fit is performed on their TOD power spectra. See \cite{lunde:2024, comap-III} for details.}
    }
\end{table*}

\FloatBarrier
\section{A simple end-to-end signal injection test}\label{apdx:signal_inj_result}
The same type of simulations used to estimate the filter transfer function (see Sect. \ref{sec:sim_pipeline}) can also be used to perform a rudimentary end-to-end signal injection test to confirm the detectability of signal given our transfer function estimate. The signal injection pipeline is explained in more detail in Sect.~6.1 of \cite{lunde:2024}, and we here limit the scope to the application thereof.

As the signal and noise in the raw data are affected by both the low-level analysis and instrumental effects described by \cite{lunde:2024}, an important question to answer is whether we would be able to reconstruct the amplitude of an amplified CO signal within the estimated uncertainties using the earlier described FGPXS method. For instance, we know that several of the PCA filters in the low-level pipeline detailed by \cite{lunde:2024} can in principle act non-linearly if the CO S/N becomes too high. One therefore has to verify that the filters remove equal amounts of signal at any given $k$-mode of the map in the filter transfer function estimation and for the actual signal estimation from the data. Otherwise, the final signal estimate computed obtained from the data would be biased.

We generate mock signal maps to use in this injection mechanism by applying the fiducial halo model to dark matter halos simulated with the peak-patch technique~\citep{bond:1996, stein:2019}. Furthermore, we use a raw COMAP data volume corresponding to all the fast-moving azimuth scans of Field 3. This roughly corresponds to the largest independently filtered data volume or, in other words, the highest possible CO S/N. We then boost the injected signal by a factor of three before injecting it into the raw time stream to ensure the signal is detectable above the instrumental noise. Subsequently, the TOD are filtered and binned into maps using the pipeline described by \cite{lunde:2024}, before we compute the FGPXS signal estimate to see whether the injected signal was successfully recovered. 
We note that only one signal realization is used because these high-realism mocks are expensive to produce. However, the test still functions as a simple qualitative ``sanity check'' that the pipeline works as intended. Future work will expand on this modest check by including further signal realizations and COMAP fields.

\begin{figure}[!ht]
    \centering
    \includegraphics[width=\linewidth]{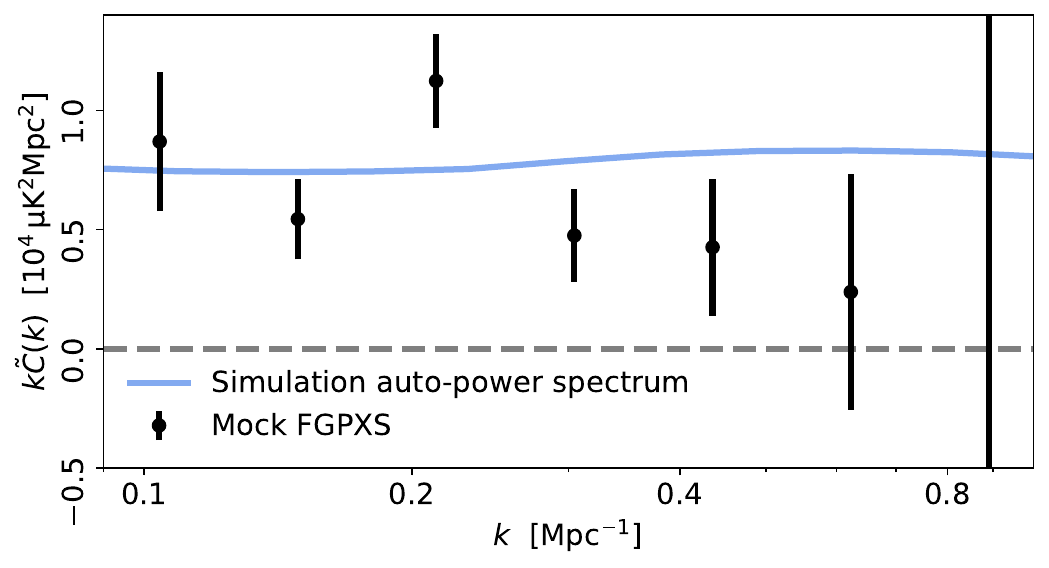}
    \caption{Example of the spherically averaged FGPXS (black points) resulting from injecting a mock CO signal realization (blue input power spectrum) of the (line-broadened) COMAP fiducial model \citep{chung:2021, comap-V} with a boost factor of three into all fast-moving azimuth data of Field 3.}
    \label{fig:pxs_end2end}
\end{figure}

The resulting mock FGPXS data points as well as the auto-power spectrum of the input simulation can be seen in Fig.~\ref{fig:pxs_end2end}. We can clearly see a high-significance excess that appears consistent with the power spectrum of the input signal within the estimated error bars (which are estimated using the RND methodology described in Sec.~\ref{subsec:rnd_errors}). The excess is large enough to place the computed $\chi^2$-value of the mock data far outside the computed RND $\chi^2$-distribution. Therefore, to assign a quantitative value to the significance of this mock detection, we instead use the simplified assumptions of approximately Gaussian uncertainties. When doing so, we obtain an estimated $\approx 6\sigma$ detection of non-zero power. Meanwhile, testing against the input signal we get a $1.5\sigma$ significance away from the model, meaning we recover the input signal within at most mild tension.

This exercise demonstrates that we can recover the input signal within the experimental errors, indicating that our pipeline, the full transfer function, and error bar estimation work as expected.
 
\end{appendix}

\end{document}